\documentclass[final,5p,times,twocolumn,authoryear]{elsarticle}

\usepackage{amssymb}
\usepackage{lipsum}
\usepackage{amsmath}	% Advanced maths commands
\usepackage{xcolor}     % Handle colours.
\usepackage{suffix}     % Easily create * version of commands.
\usepackage{algorithm}
\usepackage{algpseudocodex}
\usepackage{subcaption}
\usepackage{hyperref}
\usepackage{verbatim}

\newcommand{\quartical}{\textsc{QuartiCal}}
\newcommand{\stimela}{\textsc{stimela}}
\newcommand{\pfb}{\textsc{pfb-imaging}}
\newcommand{\cubical}{\textsc{CubiCal}}
\newcommand{\dask}{\textsc{Dask}}
\newcommand{\zarr}{\textsc{Zarr}}
\newcommand{\pydata}{\textsc{PyData}}
\newcommand{\daskms}{\textsc{Dask-MS}}
\newcommand{\codexafricanus}{\textsc{codex africanus}}
\newcommand{\numpy}{\textsc{NumPy}}
\newcommand{\numba}{\textsc{Numba}}

\newcommand{\xarray}{\textsc{XArray}}
\newcommand{\red}[1]{\textcolor{red}{#1}}

\newcommand{\REAL}{\mathbb{R}}
\newcommand{\COMPLEX}{\mathbb{C}}
\newcommand{\NATURAL}{\mathbb{N}}

\newcommand{\vecop}[2][\big]{\underset{r}{\mathrm{vec}} #1( #2 #1)}
\newcommand{\ivecop}[2][\big]{\underset{r}{\mathrm{vec}}^{-1} #1( #2 #1)}

\newcommand{\gmat}[2]{\mathbf{G}^{(#1)}_{#2}}
\WithSuffix\newcommand\gmat*[2]{\mathbf{G}^{#1}_{#2}}
\newcommand{\gmatct}[2]{\mathbf{G}^{(#1)\dagger}_{#2}}
\WithSuffix\newcommand\gmatct*[2]{\mathbf{G}^{#1\dagger}_{#2}}
\newcommand{\cmat}[2]{\mathbf{C}^{(#1)}_{#2}}
\WithSuffix\newcommand\cmat*[2]{\mathbf{C}^{#1}_{#2}}
\newcommand{\cmatct}[2]{\mathbf{C}^{(#1)\dagger}_{#2}}
\WithSuffix\newcommand\cmatct*[2]{\mathbf{C}^{#1\dagger}_{#2}}
\newcommand{\mmat}[1]{\mathbf{M}^{}_{#1}}

\newcommand{\doalign}[2]{%
 {\vbox{\offinterlineskip\ialign{\hfil##\hfil\cr#1\cr$#2$\cr}}}%
}
\newcommand{\rharp}[1]{\mathpalette\rharpoonvec{#1}}
\newcommand{\rharpvecsign}{\scriptscriptstyle\rightharpoonup}
\newcommand{\rharpoonvec}[2]{%
  \ifx\displaystyle#1\doalign{$\rharpvecsign$}{#1#2}\fi
  \ifx\textstyle#1\doalign{$\rharpvecsign$}{#1#2}\fi
  \ifx\scriptstyle#1\doalign{\scalebox{.6}[.9]{$\rharpvecsign$}}{#1#2}\fi
  \ifx\scriptscriptstyle#1\doalign{\scalebox{.5}[.8]{$\rharpvecsign$}}{#1#2}\fi
}
\newcommand{\lharp}[1]{\mathpalette\lharpoonvec{#1}}
\newcommand{\lharpvecsign}{\scriptscriptstyle\leftharpoonup}
\newcommand{\lharpoonvec}[2]{%
  \ifx\displaystyle#1\doalign{$\lharpvecsign$}{#1#2}\fi
  \ifx\textstyle#1\doalign{$\lharpvecsign$}{#1#2}\fi
  \ifx\scriptstyle#1\doalign{\scalebox{.6}[.9]{$\lharpvecsign$}}{#1#2}\fi
  \ifx\scriptscriptstyle#1\doalign{\scalebox{.5}[.8]{$\lharpvecsign$}}{#1#2}\fi
}

\newcommand{\cmatl}[2]{\lharp{\mathbf{C}}^{(#1)}_{#2}}
\WithSuffix\newcommand\cmatl*[2]{\lharp{\mathbf{C}}^{#1}_{#2}}
\newcommand{\cmatlct}[2]{\lharp{\mathbf{C}}^{(#1)\dagger}_{#2}}
\WithSuffix\newcommand\cmatlct*[2]{\lharp{\mathbf{C}}^{#1\dagger}_{#2}}

\newcommand{\cmatr}[2]{\rharp{\mathbf{C}}^{(#1)}_{#2}}
\WithSuffix\newcommand\cmatr*[2]{\rharp{\mathbf{C}}^{#1}_{#2}}
\newcommand{\cmatrct}[2]{\rharp{\mathbf{C}}^{(#1)\dagger}_{#2}}
\WithSuffix\newcommand\cmatrct*[2]{\rharp{\mathbf{C}}^{#1\dagger}_{#2}}

\newcommand{\gvec}{\mathbf{g}}
\newcommand{\gvecc}{\mathbf{\bar g}}
\newcommand{\uvec}{\mathbf{u}}
\newcommand{\rvec}{\mathbf{r}}
\newcommand{\rvecc}{\mathbf{\bar r}}
\newcommand{\dvec}{\mathbf{d}}
\newcommand{\vvec}{\mathbf{v}}
\newcommand{\vvecc}{\mathbf{\bar v}}
\newcommand{\guvec}{\uvec_{(\cdot)}}
\newcommand{\mvec}{\mathbf{m}}

\newcommand{\wmat}{\mathbf{W}}
\newcommand{\jmat}{\mathbf{J}}
\newcommand{\imat}{\mathbf{I}}
\newcommand{\pmat}{\mathbf{P}}
\newcommand{\amat}{\mathbf{A}}

\newcommand{\dmat}{\mathbf{D}}
\newcommand{\vmat}{\mathbf{V}}
\newcommand{\emat}{\mathbf{E}}

\newcommand{\jmatul}{\jmat^{\vphantom{\dagger}}_{\gvec}}
\newcommand{\jmatulct}{\jmat^{\dagger}_{\gvec}}
\newcommand{\jmatur}{\jmat^{\vphantom{\dagger}}_{\gvecc}}
\newcommand{\jmatll}{\bar{\jmat}^{\vphantom{\dagger}}_{\gvecc}}
\newcommand{\jmatllct}{\bar{\jmat}^{\dagger}_{\gvecc}}
\newcommand{\jmatlr}{\bar{\jmat}^{\vphantom{\dagger}}_{\gvec}}

\newcommand{\jpmatu}{\jmat^{\vphantom{\dagger}}_{\uvec}}
\newcommand{\jpmatl}{\bar{\jmat}^{\vphantom{\dagger}}_{\uvec}}

\newcommand{\jgumatu}{\jmat^{\vphantom{\dagger}}_{\circ}}
\newcommand{\jgumatl}{\bar{\jmat}^{\vphantom{\dagger}}_{\circ}}
\newcommand{\jgumatuct}{\jmat^{\dagger}_{\circ}}
\newcommand{\jgumatlct}{\bar{\jmat}^{\dagger}_{\circ}}

\newcommand{\nant}{N_{A}}
\newcommand{\ncor}{N_{C}}
\newcommand{\ndir}{N_{D}}

\newcommand{\nbl}{N_{bl}}
\newcommand{\nsam}{N_{s}}
\newcommand{\njay}{N_{j}}

\newcommand{\aset}{\mathcal{A}}
\newcommand{\dset}{\mathcal{D}}
\newcommand{\bset}{\mathcal{B}}
\newcommand{\sset}{\mathcal{S}}
\newcommand{\rset}{\mathcal{R}}
\newcommand{\cset}{\mathcal{C}}

\newcommand{\ahess}{\jmat^{\dagger} \wmat \jmat}
\newcommand{\jhwr}{\jmat^{\dagger} \wmat \breve{\rvec}}

\newcommand{\realop}{\mathfrak{R}}

\journal{Astronomy $\&$ Computing}

\begin{document}

\begin{frontmatter}



\title{Africanus II. \quartical{}: calibrating radio interferometer data at scale using \numba{} and \dask{}}

\author[ratt]{J.~S.~Kenyon}\ead{jonosken@gmail.com}
\author[sarao]{S.~J.~Perkins}
\author[sarao,ratt]{H.~L.~Bester}
\author[ratt,sarao,ira]{O.~M.~Smirnov}
\author[ratt]{C.~Russeeawon}
\author[sarao]{B.~V.~Hugo}

\affiliation[ratt]{organization={Centre for Radio Astronomy Techniques \& Technologies (RATT),
Department of Physics and Electronics, Rhodes University},
            city={Makhanda},
            state={EC},
            country={South Africa}}

\affiliation[sarao]{organization={South African Radio Astronomy Observatory (SARAO)},
            city={Cape Town},
            state={WC},
            country={South Africa}}

\affiliation[ira]{organization={Institute for Radioastronomy, National Institute of Astrophysics (INAF IRA)},
            city={Bologna},
            country={Italy}}

\begin{abstract}
Calibration of radio interferometer data ought to be a solved problem; it has been an integral part of data reduction for some time. However, as larger, more sensitive radio interferometers are conceived and built, the calibration problem grows in both size and difficulty.

The increasing size can be attributed to the fact that the data volume scales quadratically with the number of antennas in an array. Additionally, new instruments may have up to two orders of magnitude more channels than their predecessors. Simultaneously, increasing sensitivity is making calibration more challenging: low-level RFI and calibration artefacts (in the resulting images) which would previously have been subsumed by the noise may now limit dynamic range and, ultimately, the derived science.

It is against this backdrop that we introduce \quartical{}: a new Python package implementing radio interferometric calibration routines. \quartical{} improves upon its predecessor, \cubical{}, in terms of both flexibility and performance. Whilst the same mathematical framework - complex optimization using Wirtinger derivatives - is in use, the approach has been refined to support arbitrary length chains of parameterized gain terms.

\quartical{} utilizes \dask{}, a library for parallel computing in Python, to express calibration as an embarrassingly parallel task graph. These task graphs can (with some constraints) be mapped onto a number of different hardware configurations, allowing \quartical{} to scale from running locally on consumer hardware to a distributed, cloud-based cluster.

\quartical{}'s qualitative behaviour is demonstrated using MeerKAT observations of PSR J2009-2026. These qualitative results are followed by an analysis of \quartical{}'s performance in terms of wall time and memory footprint for a number of calibration scenarios and hardware configurations. 
\end{abstract}



\begin{keyword}
%% keywords here, in the form: keyword \sep keyword, up to a maximum of 6 keywords
Radio Astronomy \sep Calibration \sep Software \sep Distributed Computing \sep Numerical Methods



\end{keyword}

\end{frontmatter}

%\tableofcontents



\section{Introduction}

With the ubiquity of Big Data in modern data processing \citep{villars2011big}, one might expect that the calibration (correction of antenna-based gains) of a radio interferometer would not pose too great a challenge. However, it is a challenge that is perpetually growing, largely as a result of the ambitious scale and design goals of both existing and upcoming instruments e.g. LOFAR \citep{vanhaarlem2013}, MeerKAT (and ultimately the SKA) \citep{dewdney2009}, and DSA-2000 \citep{hallinan2019}. In fact, how (in practical terms) to process the data from the largest of these projects is still an open question and an area of ongoing research.

At the root of this issue is the fact that the number of baselines in an interferometer grows quadratically with the number of antennas/receiving elements \citep{whitebook2017}. This, coupled with the fact that increased frequency resolution is crucial for many science cases e.g. spectral line surveys \citep{wagenveld2023}, ensures that the size of radio interferometric data products is growing too fast for processing to keep up. A key factor in this burgeoning divide is the lack of widely-adopted calibration software with native support for distributed compute.

The most commonly used calibration software, CASA \citep{casa2007,casa2022}, was designed prior to the relatively recent explosion in data volume and before distributed/cloud computing had become the norm. While support for distribution via MPI (Message Passing Interface, \citet{mpi40}) has subsequently been added to CASA \citep{roberts1999,castro2017}, this post-hoc approach requires manual partitioning of input data and, according to the documentation\footnote{\url{https://casa.nrao.edu/casadocs/casa-6.1.0}}, is still regarded as experimental.

Using MPI to add distributed capabilities to software after the fact is not unusual and is also planned/in-progress for DDFacet \citep{tasse2018}, a state-of-the-art imaging application which is routinely used in the LOFAR Two-metre Sky Survey (LoTSS) data reduction pipeline \citep{shimwell2019}. It remains to be seen whether the addition of distributed capabilities in this fashion will be sufficient.

Conversely, it is quite rare for calibration software to be written with distribution in mind. One of the few examples of this is SAGECAL \citep{yatawatta2015}, which implements a number of different algorithms for gain calibration. This includes distributed stochastic calibration with consensus optimization \citep{yatawatta2020} which has been implemented using MPI. Theoretically, this allows it to scale to very large problem sizes but this has not been clearly demonstrated on real-world problems in the associated literature. Other distributed approaches have been suggested in e.g. \citet{ollier2018}, but these have not yet produced usable software.

It is worth noting that no widely utilized calibration software provides all the functionality required to deal with the entirety of the calibration process. This is easily explicable as it is a daunting task, including as it does \citep[see][for a more complete description]{noordam2010,smirnov2011}:
\begin{itemize}
    \item 1GC (first generation calibration): calibration of calibrator sources (sources with known flux, morphology and spectral behaviour) and subsequent solution transfer to target fields.
    \item 2GC (second generation calibration): iterative calibration of a field, consisting of sequential improvements to a source model and gain solutions.
    \item 3GC (third generation calibration): as 2GC, but with the addition of direction-dependent gain terms i.e. terms which vary across the field of view.
\end{itemize}

%\red{@me: Somehow work in the notion that increased sensitivity means that calibration is tougher due to e.g. low level rfi? Possibly cite that Wijnholds paper.}

With all of the above in mind, it is clear that there is a critical need for end-to-end, user-friendly calibration software with well-integrated distributed compute capabilities. Unfortunately, the predecessor to this work, \cubical{} \citep{kenyon2018}, cannot be adapted to this task as it is highly stateful and relies heavily on shared memory. However, the mathematics upon which it is based \citep[complex optimization using Wirtinger derivatives; see][]{smirnov2015} is well-suited to expressing calibration as an embarrassingly parallel problem which is (at least theoretically) trivial to distribute. We will briefly revisit the mathematics in $\S$~\ref{section:mathematics}, concentrating on finite length chains of gain terms. We will also highlight the insights necessary to include arbitrarily parameterized terms in these chains and the compromises required to accommodate per-correlation weights. Armed with the resulting simple and relatively easy to implement parameter/gain updates, we can consider the building blocks out of which we can ultimately construct an implementation.

$\S$~\ref{section:software} will briefly introduce the software packages which form the basis of our Python implementation. We have chosen to develop in the Python language due to the extensive support of the \pydata{} ecosystem \citep[see Paper I;][]{africanus1} and the fact that it is widely used and understood in radio astronomy. While Python is not typically regarded as a high-performance language due to limitations such as the GIL\footnote{\url{https://wiki.python.org/moin/GlobalInterpreterLock}} (Global Interpreter Lock), one can almost entirely mitigate this limitation by using \numba{} \citep{lam2015}, a JIT (just-in-time) compiler for a subset of Python and \numpy{} code. This can provide C-like speed while retaining much of Python's simplicity. In addition, due to the extensive use of distributed/cloud computing in both industry and research, there are multiple packages which provide distributed functionality within the Python ecosystem e.g. mpi4py \citep{dalcin2021}, \dask{} \citep{rocklin2015} and Ray \citep{moritz2017}. We will focus our attention on \dask{}, mainly for the reasons outlined in Paper I and in order to capitalize on the functionality of the packages (\daskms{}, \codexafricanus{}) described therein.

The implementation details of our new package, \quartical{}, will be described in $\S$~\ref{section:implementation}. \quartical{}, which started life as \cubical{} 2.0, is a Python application which implements flexible and scalable gain calibration using the packages mentioned above. \quartical{} boasts an extensive list of features which includes support for 1GC through to 3GC. New features are being added regularly, and an effort will be made to describe some of them here.

Of course, no description of a new piece of software would be complete without demonstrating that it works. To this end, $\S$~\ref{section:results} will present results on real MeerKAT data. These results will include a qualitative demonstration of \quartical{}'s ability to calibrate data in both the direction-independent and direction-dependent regimes. This will be followed by comparisons between the single node performance of both \cubical{} and \quartical{} in terms of both memory footprint and wall time. Lastly, we will produce benchmarks of \quartical{}'s performance in a distributed cloud environment (an AWS deployment) for varying compute resources.

Finally, we will present our conclusions and ruminate on the future directions in which \quartical{} may evolve.
\section{Mathematics}
\label{section:mathematics}

This section will revisit the mathematics of \citet{kenyon2018} and endeavour to present an updated version of the formalism therein. We will focus on the update equations for arbitrary length chains of gain terms \citep[Jones chains, see][]{smirnov2011} which may include parameterized terms.

\subsection{The minimization problem}

Before attempting to write down update rules for the complicated case mentioned above, it is first necessary to formulate the minimization problem. Whilst this was done in \citet{kenyon2018}, the version we will present here is more general.

At its core, calibration involves solving a (weighted) non-linear least squares (NLLS) problem of the following form:
% \begin{equation}
% \label{eq:min}
%     \min_{\gvec,\gvecc} \wmat || \rvec (\gvec, \gvecc) ||^2 \equiv \min_{\gvec,\gvecc} || \dvec - \vvec (\gvec,\gvecc)||^2,
% \end{equation}
\begin{equation}
\label{eq:min}
    \min_{\gvec,\gvecc} \rvec(\gvec,\gvecc)^\dagger \wmat \rvec (\gvec,\gvecc) \equiv \min_{\gvec,\gvecc} (\dvec - \vvec (\gvec,\gvecc))^\dagger \wmat (\dvec - \vvec (\gvec,\gvecc)),
\end{equation}
where $\rvec \in \COMPLEX^m$ is a length $m$ vector of residual values given by the difference between a data vector $\dvec \in \COMPLEX^m$ and a model vector $\vvec \in \COMPLEX^m$. Both $\vvec$ and $\rvec$ are functions of an unknown complex-valued vector $\gvec \in \COMPLEX^n$ and its conjugate $\gvecc$. $\wmat \in \REAL^{m \times m}$ is a matrix of weights which we will assume to be diagonal throughout this work. $\bar{(\cdot)}$ denotes complex conjugation, $(\cdot)^\dagger$ denotes the conjugate transpose and $(\cdot)^{-1}$ indicates a matrix inverse.

Traditional NLLS methods for solving such a problem e.g. Gauss-Newton and Levenberg-Marquardt \citep[see][for details]{madsen2004} utilize derivative information. Alas, \eqref{eq:min} describes the minimization of a real-valued function with respect to a complex-valued variable. This is problematic as the partial derivative $\partial\bar g/\partial g$ is not defined when using conventional calculus. However, the Wirtinger derivatives \citep{wirtinger1927}, allow for the derivation of complex equivalents of the Gauss-Newton and Levenberg-Marquardt algorithms \citep{kreutzdelgado2009,sorber2012}. We will focus on the complex Gauss-Newton updates as the extension to Levenberg-Marquardt is trivial.

The update rule for complex Gauss-Newton at iteration $k+1$ is given by:
\begin{equation}
\label{eq:gaussnewton}
    \breve{\gvec}_{k+1} = \breve{\gvec}_{k} - (\jmat^\dagger \wmat \jmat)^{-1} \jmat^\dagger \wmat \breve{\rvec}(\breve{\gvec}_k),
\end{equation}
where previously defined symbols retain their meanings. $\breve{(\cdot)}$ denotes complex augmentation which is the stacking of a column vector with its complex conjugate e.g:
\begin{equation}
\label{eq:augmentation}
    \breve{\rvec} =
    \begin{bmatrix}
        \rvec \\
        \bar{\rvec}
    \end{bmatrix}.
\end{equation}

As a result of this augmentation, the size of the matrix $\wmat$ increases such that $\wmat \in \REAL^{2m \times 2m}$ and each value in $\rvecc$ will be assigned the same weight as its associated value in $\rvec$.

Finally, $\jmat \in \COMPLEX^{2m \times 2n}$ is the Jacobian (matrix of first-order partial derivatives) and is given in block form by:
\begin{equation}
\label{eq:jacobian}
    \jmat = \frac{\partial \breve{\rvec}}{\partial \breve{\gvec}} =
    \renewcommand{\arraystretch}{1.4}
    % \cellspacetoplimit 3pt
    % \cellspacebottomlimit 3pt
    \setlength{\arraycolsep}{4pt}
    \begin{bmatrix}
        \frac{\partial \rvec}{\partial \gvec} &
        \frac{\partial \rvec}{\partial \gvecc} \\
        \frac{\partial \rvecc}{\partial \gvec} &
        \frac{\partial \rvecc}{\partial \gvecc}
    \end{bmatrix} =
    % \begin{bmatrix}
    %     \amat & \bmat \\
    %     \bar{\bmat} & \bar{\amat}
    % \end{bmatrix} =
    \begin{bmatrix}
        \jmatul & \jmatur \\
        \jmatll & \jmatlr
    \end{bmatrix},
\end{equation}
where $\jmatul,\jmatur \in \COMPLEX^{m \times n}$ and the final equivalence is taken from \citet{smirnov2015}. It is worth emphasizing that $\jmatlr$ is simply the complex conjugate of $\jmatul$, as $\jmatll$ is the complex conjugate of $\jmatur$. This is sufficient information to begin writing down update rules for the complex unknowns $\gvec$ and $\gvecc$. However, it is useful to first clarify how this applies to radio interferometry and to refine our definition of the model, $\vvec$.

\subsection{A flexible model for non-linear least-squares}

In the context of radio interferometry, the previously defined $\dvec$ and $\vvec$ are complex-valued visibility vectors corresponding to the observed data and our model thereof. Each of these vectors contains $\nbl \nsam \ncor$ elements, where $\nbl$ is the number of baselines (unique pairs of antennas) in the array, $\nsam$ is the total number of time and frequency samples and $\ncor$ is the number of correlations (typically four, although situations in which $\ncor < 4$ can be handled by introducing zeros). Where relevant, baselines will be indexed by $pq$, where $p$ is the index of the first antenna and $q$ is the index of the second. Baselines will typically be ordered such that $q>p$, and will not include conjugate baselines ($q<p$) or autocorrelations ($q = p$) unless otherwise stated.

The model, $\vvec$, is a function of some unknown, complex-valued column vector $\gvec \in \COMPLEX^{\nant\ndir\ncor}$ and its conjugate $\gvecc$, where $\nant$ is the number of antennas in the array, $\ndir$ is the number of directions in which the gain is to be solved and $\ncor$ is as described above. Typically, these model the per-antenna gains which describe changes in the signal. We will formally define the component of $\gvec$ associated with antenna $p$ in direction $d$ as
\begin{equation}
\label{eq:gvec}
    \gvec_{p,d} =
    \begin{bmatrix}
        g_{p,d}^{XX} & g_{p,d}^{XY} & g_{p,d}^{YX} & g_{p,d}^{YY}
    \end{bmatrix}^T,
\end{equation}
and note that $\gvec$ consists of $\nant\ndir$ such components stacked on top of each other. $(\cdot)^T$ denotes transposition as normal.

It is often convenient to express these components of the gain vector as $2 \times 2$ Jones matrices \citep{hamaker1996}
\begin{equation}
\label{eq:jones}
    \mathbf{G}_{p,d} =
        \renewcommand{\arraystretch}{1.3}
        % \cellspacetoplimit 3pt
        % \cellspacebottomlimit 3pt
        \setlength{\arraycolsep}{4pt}
        \begin{bmatrix}
            g_{p,d}^{XX} & g_{p,d}^{XY} \\ g_{p,d}^{YX} & g_{p,d}^{YY}
        \end{bmatrix}
    \equiv
    \ivecop{\gvec_{p,d}},
\end{equation}
where $\ivecop{\cdot}$ denotes the inverse of row-major vectorization, accomplished by unstacking the entries of column vector $\gvec_{p,d}$ as shown above. We choose to consider row-major vectorization as it is consistent with both the measurement set, one of the most prevalent data formats in radio interferometry, and the array ordering of the C programming language. In order to simplify later derivations, we also draw the readers' attention to the following relationship:
\begin{equation}
\label{eq:jonesconj}
    \vecop{\mathbf{G}^\dagger_{p,d}} =
        \renewcommand{\arraystretch}{1.3}
        % \cellspacetoplimit 3pt
        % \cellspacebottomlimit 3pt
        \setlength{\arraycolsep}{4pt}
        \begin{bmatrix}
            \bar{g}_{p,d}^{XX} & \bar{g}_{p,d}^{YX} & \bar{g}_{p,d}^{XY} & \bar{g}_{p,d}^{YY}
        \end{bmatrix}^T
    \equiv
    \mathbf{P}\gvecc_{p,d},
\end{equation}
where $\pmat$ is an involutory permutation matrix given by:
\begin{equation}
\label{eq:permutation}
    \pmat =
    \pmat^{-1} =
    \begin{bmatrix}
        1 & 0 & 0 & 0 \\
        0 & 0 & 1 & 0 \\
        0 & 1 & 0 & 0 \\
        0 & 0 & 0 & 1
    \end{bmatrix}.
\end{equation}
The preceding expressions have been expressed in terms of linear feeds but equivalent expressions can be obtained for circular feeds by replacing $X$ with $R$ and $Y$ with $L$.

Solving for a single gain or Jones term is relatively simple and is performed regularly during data reduction. However, it is often desirable to disentangle the various sources of error by solving for a Jones chain \citep[see][]{smirnov2011}, which includes one or more Jones/gain terms, each describing a change in the signal as it propagates. Each of these terms may be defined over a different region of contiguous input samples (commonly referred to as a solution interval) and in one or more directions. Jones terms which vary with direction can be used to model direction-dependent (3GC) effects. We will assume that all terms are direction-dependent in the following derivations; this helps simplify the notation without loss of generality as a direction-independent term can be regarded as a direction-dependent term which is the same in all directions. We will return to this point later.

Note that while we will hereafter consider Jones chains of arbitrary length, we will only ever consider updating a single term in the chain at a time - the alternative is computationally impractical and may exhibit degeneracies.

Let us formally define a Jones chain associated with antenna $p$ in direction $d$ as follows:
\begin{equation}
\label{eq:joneschain}
     \cmat*{}{p,d} = \prod_{j=1}^{\njay} \gmat*{}{j,p,d},
\end{equation}
where $j$ indexes an arbitrary number, $\njay$, Jones terms. From the above, it is trivial to demonstrate that:
\begin{equation}
\label{eq:joneschainct}
     \cmatct*{}{p,d}
     = \Big( \prod_{j=1}^{\njay} \gmat*{}{j,p,d} \Big)^\dagger
     = \prod_{j=N_j}^{1} \gmatct*{}{j,p,d}.
\end{equation}
For reasons that will soon become apparent, it is convenient to define the following alternative factorization:
\begin{equation}
\label{eq:jonesfactor}
    \begin{aligned}[b]
        \cmat*{}{p,d}
        &= \Big( \prod_{j=1}^{n-1} \gmat*{}{j,p,d} \Big) \gmat*{}{n,p,d} \Big( \prod_{j=n+1}^{\njay} \gmat*{}{j,p,d} \Big) \\
        &= \cmatl*{}{n,p,d} \gmat*{}{n,p,d} \cmatr*{}{n,p,d},
    \end{aligned}
\end{equation}
where $\cmatl*{}{n,p,d}$ and $\cmatr*{}{n,p,d}$ can be thought of intuitively as the components of chain $\cmat*{}{p.d}$ to the left and right of the $n$-th term respectively. In situations where $\njay = 1$,  $\cmatl*{}{n,p,d}$ and $\cmatr*{}{n,p,d}$ can be replaced with the identity matrix, $\imat \in \REAL^{2\times2}$. This is also true for cases where $n$ is either the first or last element in the chain; the empty side of the chain can be treated as the identity matrix.

Using \eqref{eq:joneschain} and \eqref{eq:joneschainct}, we can write down the following expression for the model associated with a specific baseline and sample:
\begin{equation}
\label{eq:baselinemodel}
    \vvec_{pq,s}
    =
    \vecop{\mathbf{V}_{pq,s}}
    =
    \vecop{\sum_{d=1}^{\ndir} \cmat*{}{p,d} \mmat{pq,s,d} \cmatct*{}{q,d}},
\end{equation}
where $d$ indexes an arbitrary positive integer number of directions, $\ndir$, $s$ indexes the time and frequency samples and $\vecop{\cdot}$ denotes vectorization by row stacking. $\mmat{pq,s,d} \in \COMPLEX^{2 \times 2}$ is the \textit{predicted visibility} Jones matrix associated with baseline $pq$ in direction $d$ for sample index $s$ and should not be confused with the \textit{model visibility} Jones matrix $\mathbf{V}_{pq,s} \in \COMPLEX^{2 \times 2}$, which includes the contributions of both the predicted visibility and all other Jones terms. The complete model, $\vvec$, is then made up of $\nbl\nsam$ such vectors stacked on top of each other.

Equation \eqref{eq:baselinemodel}, which is simply a generic form of the radio interferometer measurement equation \citep[or RIME, see][]{hamaker1996,smirnov2011}, is the major result of this section: an expression for the model in terms of an arbitrary number of potentially direction-dependent antenna-based Jones terms. This can be used to represent a wide variety of complicated cases.

\subsection{Deriving an update rule}

Armed with \eqref{eq:baselinemodel}, we are in a position to return to the derivation of the Jacobian. This matrix is one of the fundamental building blocks on which the NLLS methods under discussion rely and it behooves us to construct a suitable generic expression for it.

It is clear from \eqref{eq:jacobian} that we require expressions for the elements of the matrices $\jmatul$ and $\jmatur$. In order to write down these expressions we first need to combine \eqref{eq:baselinemodel} with our definition of the residual:
\begin{equation}
\label{eq:baselineresidual}
    \begin{aligned}[b]
        \rvec_{pq,s}
        & = \dvec_{pq,s} - \vvec_{pq,s} \\
        & = \vecop{\dmat_{pq,s}} - \vecop{\vmat_{pq,s}} \\
        & = \vecop{\dmat_{pq,s}} - \vecop{\sum_{d=1}^{\ndir} \cmat*{}{p,d} \mmat{pq,s,d} \cmatct*{}{q,d}},
    \end{aligned}
\end{equation}
where $\dvec_{pq,s} \in \COMPLEX^{\ncor}$ is the column vector of observed data associated with baseline $pq$ and sample $s$ and $\mathbf{D}_{pq,s} \in \COMPLEX^{2 \times 2}$ is its Jones matrix equivalent. The residual vector over all baselines, samples and correlations $\rvec \in \COMPLEX^{\nbl\nsam\ncor}$ consists of $\nbl\nsam$ such column vectors stacked on top of each other.

Differentiating \eqref{eq:baselineresidual} with respect to the gain vector is possible but can be simplified using the row-major version of the so-called ``vec trick'' \citep{roth1934}. Given the matrix equation $\mathbf{C} = \mathbf{A} \mathbf{X} \mathbf{B}$, we can use this trick to rewrite it in the following form:
\begin{equation}
\label{eq:vectrick}
    \vecop{\mathbf{C}} = \vecop{\mathbf{A} \mathbf{X} \mathbf{B}} = (\mathbf{A} \otimes \mathbf{B}^T) \vecop{\mathbf{X}},
\end{equation}
where $\otimes$ denotes the Kronecker product and $\mathbf{A}$, $\mathbf{X}$ and $\mathbf{B}$ are arbitrary compatible matrices. This is useful in that it allows us to rewrite \eqref{eq:baselinemodel} such that the gain with respect to which we are differentiating corresponds to $\mathbf{X}$ in the ``vec trick'' expressions. This, in turn, makes the required vector-by-vector derivatives trivial.

It is perhaps simplest to demonstrate this with an example. Let us consider \eqref{eq:baselineresidual} and assume that we ultimately wish to take the derivative with respect to $\gvec_{n,p,d}$, the vector form of the $j=n$ gain in chain $\cmat*{}{p,d}$. We can express this simply as:
\begin{equation}
\label{eq:residual_derivative}
    \begin{aligned}[b]
        \frac{\partial \rvec_{pq,s}}{\partial \gvec_{n,p,d}}
        & = \frac{\partial \dvec_{pq,s}}{\partial \gvec_{n,p,d}} - \frac{\partial \vvec_{pq,s}}{\partial \gvec_{n,p,d}}.
    \end{aligned}
\end{equation}

The first term in this expression is always zero as the observed data has no dependence on our gain estimates. It is the second term which is of interest and to which we will apply the ``vec trick''. Starting with the model $\vvec_{pq,s}$ and recalling the factorization in \eqref{eq:jonesfactor}, we have:
\begin{equation}
\label{eq:chain_trick}
    \begin{aligned}[b]
        \vvec_{pq,s}
        & = \vecop[\bigg]{\sum_{d=1}^{\ndir} \cmat*{}{p,d} \mmat{pq,s,d} \cmatct*{}{q,d}} \\
        & = \vecop[\bigg]{\sum_{d=1}^{\ndir} \cmatl*{}{n,p,d} \gmat*{}{n,p,d} \cmatr*{}{n,p,d} \mmat{pq,s,d} \cmatct*{}{q,d}} \\
        & = \sum_{d=1}^{\ndir} \bigg( \cmatl*{}{n,p,d} \otimes \big(\cmatr*{}{n,p,d} \mmat{pq,s,d} \cmatct*{}{q,d}\big)^T \bigg) \vecop{\gmat*{}{n,p,d}} \\
        & = \sum_{d=1}^{\ndir} \bigg( \cmatl*{}{n,p,d} \otimes \big(\cmatr*{}{n,p,d} \mmat{pq,s,d} \cmatct*{}{q,d}\big)^T \bigg) \gvec_{n,p,d},
    \end{aligned}
\end{equation}
where the final modification exploits the fact that $\vecop{\gmat*{}{n,p,d}} = \gvec_{n,p,d} $. From standard numerator layout matrix calculus, we know that $\frac{\partial \amat\mathbf{x}}{\partial \mathbf{x}} = \amat$. However, we first need to make a distinction between the direction-independent and direction-dependent cases. In the former, $\gvec_{n,p,d} = \gvec_{n,p}$ and it becomes a common factor in all the terms of the summation. Consequently, we can pull it out on the right-hand side and the derivative is given by:
\begin{equation}
\label{eq:g_derivative_di}
    \begin{aligned}[b]
        \frac{\partial \vvec_{pq,s}}{\partial \gvec_{n,p}}
        & = \sum_{d=1}^{\ndir} \bigg( \cmatl*{}{n,p,d} \otimes \big(\cmatr*{}{n,p,d} \mmat{pq,s,d} \cmatct*{}{q,d}\big)^T \bigg).
    \end{aligned}
\end{equation}
% One peculiarity of the above expression is that the summation over direction remains even though we took the derivative with respect to a direction-independent gain. This reflects the fact that other terms in the model may be direction-dependent and it is necessary to take the sum

In the latter, direction-dependent case, differentiation acts as a form of selection and the derivative is given by:
\begin{equation}
\label{eq:g_derivative_dd}
    \begin{aligned}[b]
        \frac{\partial \vvec_{pq,s}}{\partial \gvec_{n,p,d}}
        & = \cmatl*{}{n,p,d} \otimes \big(\cmatr*{}{n,p,d} \mmat{pq,s,d} \cmatct*{}{q,d}\big)^T.
    \end{aligned}
\end{equation}
A similar procedure can be applied to the case where we wish to differentiate with respect to $\gvecc_{n,q,d}$, yielding
\begin{equation}
\label{eq:gct_derivative_di}
    \begin{aligned}[b]
        \frac{\partial \vvec_{pq,s}}{\partial \gvecc_{n,q}}
        & = \sum_{d=1}^{\ndir} \Big( \big( \cmat*{}{p,d} \mmat{pq,s,d} \cmatrct*{}{n,q,d} \big)  \otimes \big( \cmatlct*{}{n,q,d} \big)^T \Big) \pmat,
    \end{aligned}
\end{equation}
in the direction-independent case and
\begin{equation}
\label{eq:gct_derivative_dd}
    \begin{aligned}[b]
        \frac{\partial \vvec_{pq,s}}{\partial \gvecc_{n,q,d}}
        & = \Big( \big( \cmat*{}{p,d} \mmat{pq,s,d} \cmatrct*{}{n,q,d} \big)  \otimes \big( \cmatlct*{}{n,q,d} \big)^T \Big) \pmat,
    \end{aligned}
\end{equation}
in the direction-dependent case, where the permutation matrix $\pmat$ appears as a result of applying \eqref{eq:jonesconj}.

In \eqref{eq:g_derivative_di} through \eqref{eq:gct_derivative_dd} we have only considered differentiating with respect to gain terms which appear in the expression for the model visibility i.e. $\gvec_{n,p,d}$ and $\gvecc_{n,q,d}$ both appear in $\vvec_{pq,s}$. To derive expressions for the Jacobian, we need to consider the more general case in which the gain term with respect to which we are differentiating may not appear in a particular model visibility. Fortunately, this poses no difficulty as in all cases where the gain of interest does not appear,  the derivative is the $4 \times 4$ zero matrix, $\mathbf{0}_{4 \times 4}$.

With these derivative expressions safely in hand, we are in a position to return to \eqref{eq:jacobian} and write down expressions for $\jmatul$ and $\jmatur$. This is simplest if we regard each entry in these matrices as a $4 \times 4$ block given by the above derivatives. However, it is still difficult to associate a specific matrix element with the appropriate derivative. This is largely due to the fact that the indices of a two-dimensional matrix do not map directly to the indices of our equations (indexed by baseline and sample) and unknowns (indexed by antenna and direction). To remedy this situation, we define the following sets:
\begin{align}
\label{eq:blset}
    \sset &= \{ s \mid s \in \mathbb{N} \ \mathrm{and} \ 1 \leq s \leq \nsam \} \\
    \bset &= \{ pq \mid p,q \in \NATURAL \ \mathrm{and} \ 1 \leq p < q \leq \nant \} \\
    \rset &= \bset \times \sset = \{ (pq,s) \mid pq \in \bset \ \mathrm{and} \ s \in \sset \} \\
    \aset &= \{ a \mid a \in \mathbb{N} \ \mathrm{and} \ 1 \leq a \leq \nant \} \\
    \dset &= \{ d \mid d \in \mathbb{N} \ \mathrm{and} \ 1 \leq d \leq \ndir \} \\
    \cset &= \aset \times \dset = \{ (a,d) \mid a \in \aset \ \mathrm{and} \ d \in \dset \}
\end{align}
where $\NATURAL$ is the set of all natural numbers and $\rset_i$ denotes the $i$-th element of the associated set. Note that these definitions are appropriate for the direction-dependent case; for the direction-independent case we simply omit $d$ from the definition of $\cset$.

Using $\rset$ and $\cset$ (where the notation has been chosen to suggest row and column respectively), we now have a convenient way of associating elements in these sets with particular matrix indices i.e. given row index $x$ and column index $y$, the $4 \times 4$ block at location $xy$ is given by
\begin{equation}
\label{eq:jacobian_A}
    [\jmatul]^{\vphantom{\dagger}}_{xy}
    = \frac{\partial \rvec^{\vphantom{\dagger}}_{\rset_x}}{\partial \gvec^{\vphantom{\dagger}}_{n,\cset_y}}
    = -\frac{\partial \vvec^{\vphantom{\dagger}}_{\rset_x}}{\partial \gvec^{\vphantom{\dagger}}_{n,\cset_y}}.
\end{equation}
The negative sign in the above expression comes from \eqref{eq:residual_derivative}.

The entries of the matrix $\jmatur$ are very similar and are given by:
\begin{equation}
\label{eq:jacobian_B}
    [\jmatur]^{\vphantom{\dagger}}_{xy}
    = \frac{\partial \rvec^{\vphantom{\dagger}}_{\rset_x}}{\partial \gvecc^{\vphantom{\dagger}}_{n,\cset_y}}
    = -\frac{\partial \vvec^{\vphantom{\dagger}}_{\rset_x}}{\partial \gvecc^{\vphantom{\dagger}}_{n,\cset_y}}.
\end{equation}
The remaining blocks of the Jacobian can be derived by taking the complex conjugate of $\jmatul$ and $\jmatur$.

At this juncture, we already have all the expressions required to implement the Gauss-Newton update rule given in \eqref{eq:gaussnewton}. However, using an unmodified version of this update rule is computationally expensive due to the large matrix-by-matrix products and the presence of the matrix inverse. Whilst still computationally tractable, it is preferable to consider approximations which can be made to reduce the per-iteration computational complexity by trading off per-iteration update accuracy. Several such approximations were presented in \citet{smirnov2015}.

% Is this sufficient information to justify the approximations?

We will consider the most extreme of the presented approximations, \textit{AllJones}, which assumes that we can discard all off-diagonal entries from $\ahess$. This is equivalent to assuming that there is no covariance between any of the parameters, including the gains $\gvec$ and their conjugates $\gvecc$. A consequence of this approximation is that we need only consider updating $\gvec$ which in turn means we only need to consider the left half of $\jmat$.

In light of these approximations, it is worth mentioning why we choose to compute the elements of the Jacobian analytically. Auto-differentiation, as implemented in e.g. \textsc{JAX}\footnote{\url{https://github.com/google/jax}} \citep{frostig2018}, can trivialize the construction of the Jacobian. However, this approach cannot take full advantage of the properties of the problem i.e. it will not be able to exploit the same approximations as the analytic approach. This may have a very meaningful impact on both performance and memory footprint in a real application. 

We can now apply a measure of intuition to arrive at an expression for the diagonal entries of $\ahess$. Each diagonal entry is associated with a specific element of $\cset$ and is given by the product of a row of $\jmat^{\dagger}$ with a column of $\jmat$ (and, optionally, a diagonal matrix of weights). Combining \eqref{eq:jacobian_A} with our knowledge of the structure of $\jmat$, we can write the following:
\begin{equation}
\label{eq:jhj_entry}
    \begin{aligned}[b]
        [\ahess]_{n,\cset_i}
        &= [\jmatulct]^{\vphantom{\dagger}}_{ix} \wmat [\jmatul]^{\vphantom{\dagger}}_{xi} + [\jmatllct]^{\vphantom{\dagger}}_{ix} \wmat [\jmatll]^{\vphantom{\dagger}}_{xi} \\
        &= ([\jmatul]^{\vphantom{\dagger}}_{xi})^{\dagger} \wmat [\jmatul]^{\vphantom{\dagger}}_{xi} + ([\jmatll]^{\vphantom{\dagger}}_{xi})^{\dagger} \wmat [\jmatll]^{\vphantom{\dagger}}_{xi} \\
        &=
        \sum_{x} \bigg(\frac{\partial \vvec^{\vphantom{\dagger}}_{\rset_x}}{\partial \gvec^{\vphantom{\dagger}}_{n,\cset_i}} \bigg)^\dagger \wmat_{\rset_x} \frac{\partial \vvec^{\vphantom{\dagger}}_{\rset_x}}{\partial \gvec^{\vphantom{\dagger}}_{n,\cset_i}}
        +  \\
        & \hphantom{{}=}
        \sum_{x} \bigg(\frac{\partial \vvecc^{\vphantom{\dagger}}_{\rset_x}}{\partial \gvec^{\vphantom{\dagger}}_{n,\cset_i}} \bigg)^\dagger \wmat_{\rset_x} \frac{\partial \vvecc^{\vphantom{\dagger}}_{\rset_x}}{\partial \gvec^{\vphantom{\dagger}}_{n,\cset_i}}.
    \end{aligned}
\end{equation}
Whilst this expression is valid, we can substantially improve it by making some observations regarding the problem in question and moving away from the explicit matrix form.

Firstly, we know that the derivative is zero at all locations where the gain $\gvec^{\vphantom{\dagger}}_{n,\cset_i}$ does not appear in $\vvec^{\vphantom{\dagger}}_{\rset_x}$ i.e. we can reduce our summation from considering all values of $x$ to considering only the baselines which include $\gvec^{\vphantom{\dagger}}_{n,\cset_i}$. Secondly, the two terms of the above expression are identical up to the conjugation of the model. We can exploit the fact $\vvec_{qp} = \pmat\vvecc_{pq}$ to combine them provided we change the summation to include conjugate baselines. Applying these observations and recalling that $a,d \in \cset$ we can write:
\begin{equation}
\label{eq:jhj_entry_simplified}
    \begin{aligned}[b]
        [\ahess]_{n,a,d} %{a,d}
        &=
        \sum_{q \neq a,s} \bigg(\frac{\partial \vvec^{\vphantom{\dagger}}_{aq,s}}{\partial \gvec^{\vphantom{\dagger}}_{n,a,d}} \bigg)^\dagger \wmat_{aq,s} \frac{\partial \vvec^{\vphantom{\dagger}}_{aq,s}}{\partial \gvec^{\vphantom{\dagger}}_{n,a,d}},
    \end{aligned}
\end{equation}
where the summation is over all baselines and samples including antenna $a$, excluding autocorrelations.

In the event that the gain in question is direction-independent, we are free to omit the direction index in \eqref{eq:jhj_entry_simplified}, but remind the reader that a summation over direction may still appear inside the derivative terms.

We can write down a similar expression for $\jhwr$, given by:
\begin{equation}
\label{eq:jhr_entry}
    \begin{aligned}[b]
        [\jhwr]_{n,\cset_i}
        &= [\jmatulct]^{\vphantom{\dagger}}_{ix} \wmat \rvec + [\jmatllct]^{\vphantom{\dagger}}_{ix} \wmat \rvecc \\
        &= ([\jmatul]^{\vphantom{\dagger}}_{xi})^{\dagger} \wmat \rvec + ([\jmatll]^{\vphantom{\dagger}}_{xi})^{\dagger} \wmat \rvecc \\
        &=
        \sum_{x} \bigg(\frac{\partial \vvec^{\vphantom{\dagger}}_{\rset_x}}{\partial \gvec^{\vphantom{\dagger}}_{n,\cset_i}} \bigg)^\dagger \wmat^{\vphantom{\dagger}}_{\rset_x} \rvec^{\vphantom{\dagger}}_{\rset_x}
        +
        \sum_{x} \bigg(\frac{\partial \vvecc^{\vphantom{\dagger}}_{\rset_x}}{\partial \gvec^{\vphantom{\dagger}}_{n,\cset_i}} \bigg)^\dagger \wmat^{\vphantom{\dagger}}_{\rset_x} \rvecc^{\vphantom{\dagger}}_{\rset_x}.
    \end{aligned}
\end{equation}
As in the case for $\ahess$, we can simplify this expression by combining the summations and departing from the explicit matrix form:
\begin{equation}
\label{eq:jhr_entry_simplified}
    \begin{aligned}[b]
        [\jhwr]_{n,a,d} &=
        \sum_{q \neq a,s} \bigg(\frac{\partial \vvec^{\vphantom{\dagger}}_{aq,s}}{\partial \gvec^{\vphantom{\dagger}}_{n,a,d}} \bigg)^\dagger \wmat_{aq,s} \rvec_{aq,s}.
    \end{aligned}
\end{equation}

\eqref{eq:jhj_entry_simplified} and \eqref{eq:jhr_entry_simplified} express the components of the Gauss-Newton update rule with the diagonal approximation applied in a way that does not require the explicit construction of any large matrix. We can further combine them to write the per-element update rule as:
\begin{equation}
\label{eq:update_set}
    \begin{aligned}[b]
        \gvec_{k+1,n,\cset_i}  &= \gvec_{k,n,\cset_i} - [\ahess]^{-1}_{n,\cset_i} [\jhwr]^{\vphantom{-1}}_{n,\cset_i},
    \end{aligned}
\end{equation}
or alternatively,
\begin{equation}
\label{eq:update}
    \begin{aligned}[b]
        \gvec_{k+1,n,a,d}  &= \gvec_{k,n,a,d} - [\ahess]^{-1}_{n,a,d} [\jhwr]^{\vphantom{-1}}_{n,a,d}.
    \end{aligned}
\end{equation}

We have foregone substituting in the relevant expressions (\eqref{eq:jhj_entry_simplified}, \eqref{eq:jhr_entry_simplified}, and then either \eqref{eq:g_derivative_di} or \eqref{eq:g_derivative_dd}) for the sake of brevity - the resulting expressions are simple but lengthy. These results are equivalent to those derived using operator calculus originally presented in \cite{smirnov2015} and subsequently refined in \cite{kenyon2018}. However, this form of these expressions makes certain operations easier, allowing us to relax some assumptions made in \citet{kenyon2018}. Specifically, these expressions allow us to use the per-correlation weights typically present in interferometer data.

Finally, one detail conspicuous by its absence is an explicit description of solution intervals. Conveniently, \eqref{eq:jhj_entry_simplified} and \eqref{eq:jhr_entry_simplified} already incorporate this functionality in the summation over sample i.e. we solve for a constant gain over some number of input samples. The number of samples which we include in each solution is arbitrary and, given the above formulation, we can solve each solution interval entirely independently of its neighbours. It is this property that makes it possible to implement calibration in an embarrassingly parallel fashion.

\subsection{The parameterized case}

There are situations in which it is interesting to consider parameterizing a gain term i.e. to treat the complex-valued gain as a function of other parameters. As an example, consider residual delay errors. These delay errors manifest as a per-antenna slope in phase as a function of frequency. Consequently, the gain terms can be parameterized as a function of that slope to increase SNR and reduce the degrees of freedom. For the purposes of this section, we will only consider parameterizing the gains with real values as this accounts for the vast majority of use cases.

While this is not the first time that this idea has been presented \citep[see][]{kenyon2018}, here we will generalize our previous results to the case where any term in a Jones chain may be parameterized. The parameterized equivalent of \eqref{eq:gaussnewton} (the Gauss-Newton update) is given by:
\begin{equation}
\label{eq:gaussnewton_param}
    \uvec_{k+1} = \uvec_{k} - (\jmat^\dagger \wmat \jmat)^{-1} \jmat^\dagger \wmat \breve{\rvec}(\breve{\gvec}_k(\uvec_k)),
\end{equation}
where $\uvec$ is a vector of parameters which we will treat as arbitrary for now.

In order to write down the Jacobian for the parameterized case, it is clear that we need to consider not only the derivatives of the residual with respect to the gains, ${\partial \breve{\rvec}}/{\partial \breve{\gvec}}$, but the derivatives of the gains themselves with respect to the parameters, ${\partial \breve{\gvec}}/{\partial \uvec}$. Note that $\uvec$ is not augmented as it is real-valued.
% where $\uvec \in \REAL^{\nant\ndir\ncor\npar}$ is a vector of parameters and $\npar$ is the number of parameters associated with each antenna, direction and correlation. $\npar$ is special in that it may take non-integer values e.g. if a parameterization doesn't vary with correlation, $\npar$ may include a factor of $\ncor^{-1}$, effectively reducing the length of $\uvec$.

% Let us formally define an element of this vector (for antenna $p$ in direction $d$) as:
% \begin{equation}
% \label{eq:uvec}
%     \uvec_{p,d} =
%     \begin{bmatrix}
%         \uvec_{p,d}^{XX} & \uvec_{p,d}^{XY} & \uvec_{p,d}^{YX} & \uvec_{p,d}^{YY}
%     \end{bmatrix}^T,
% \end{equation}

Fortunately, the chain rule of differentiation is applicable here i.e. we can determine the Jacobian of the parameterized problem by taking the product of the Jacobian for the complex problem (hereafter $\jmat_1$) and a second Jacobian given by:
\begin{equation}
\label{eq:param_jacobian}
    \jmat_2 = \frac{\partial \breve{\gvec}}{\partial \uvec} =
    \renewcommand{\arraystretch}{1.4}
    % \cellspacetoplimit 3pt
    % \cellspacebottomlimit 3pt
    \setlength{\arraycolsep}{4pt}
    \begin{bmatrix}
        \frac{\partial \gvec}{\partial \uvec} \\
        \frac{\partial \gvecc}{\partial \uvec}
    \end{bmatrix} =
    \begin{bmatrix}
        \jpmatu \\
        \jpmatl
    \end{bmatrix}.
\end{equation}
Taking the product of this Jacobian with that presented in \eqref{eq:jacobian} gives:
\begin{equation}
\label{eq:cp_jacobian}
    \jmat = \jmat_{1} \jmat_{2} =
    \renewcommand{\arraystretch}{1.4}
    % \cellspacetoplimit 3pt
    % \cellspacebottomlimit 3pt
    \setlength{\arraycolsep}{4pt}
    \begin{bmatrix}
        \jmatul & \jmatur \\
        \jmatll & \jmatlr
    \end{bmatrix}
    \begin{bmatrix}
        \jpmatu \\
        \jpmatl
    \end{bmatrix}
    =
    \begin{bmatrix}
        \jmatul \jpmatu + \jmatur \jpmatl \\
        \jmatll \jpmatu + \jmatlr \jpmatl
    \end{bmatrix}
    =
    \begin{bmatrix}
        \jgumatu \\
        \jgumatl
    \end{bmatrix}
    .
\end{equation}

This is a powerful, generic expression for the parameterized Jacobian. That said, its generality makes writing down simple update rules difficult without first specifying the structure of $\uvec$. This is due to the fact that the structure of $\jmat_2$ is governed by the contents of $\uvec$.

In order to proceed without specifying the exact structure of $\uvec$, we will introduce a new symbol, $\guvec$, which denotes a single, vector-valued element of $\uvec$. The exact contents of $\guvec$ remain arbitrary, and the only assumption we make is that $\uvec$ consists of some number of these component vectors stacked on top of each other. Consequently, the derivative of an element of the gain vector with respect to an element of the parameter vector can be written as $\frac{\partial \gvec_{n,s,p,d}}{\partial \guvec} \in \COMPLEX^{4 \times N}$, where $N$ is governed by the length of $\guvec$ and the additional $s$ subscript in the numerator draws attention to the fact that the parameterization may exist at a different temporal and/or spectral resolution to the gains.

In order to write down expressions similar to those given in \eqref{eq:jacobian_A} and \eqref{eq:jacobian_B}, we need to define some additional sets,
\begin{align}
\label{eq:j2_sets}
    \rset^{\prime} &= \sset \times \aset \times \dset = \{ (s,a,d) \mid s \in \sset, \ a \in \aset \ \mathrm{and} \ d \in \dset\},
\end{align}
and $\cset^{\prime}$, where we leave $\cset^{\prime}$ as arbitrary but note that it allows us to map the elements of parameter vector $\uvec$ to the columns of matrices $\jpmatu$ and $\jpmatl$. We will discuss specific choices of the elements of $\cset^{\prime}$ towards the end of this section.

It is now possible to define the elements of the aforementioned matrices in a generic way. For row index $y$ and column index $z$, we have
\begin{equation}
\label{eq:pjacobian_A}
    [\jpmatu]^{\vphantom{\dagger}}_{yz}
    = \frac{\partial \gvec^{\vphantom{\dagger}}_{n,\rset^{\prime}_y}}{\partial \uvec^{\vphantom{\dagger}}_{n,\cset^{\prime}_z}},
\end{equation}
and
\begin{equation}
\label{eq:pjacobian_B}
    [\jpmatl]^{\vphantom{\dagger}}_{yz}
    = \frac{\partial \gvecc^{\vphantom{\dagger}}_{n,\rset^{\prime}_y}}{\partial \uvec^{\vphantom{\dagger}}_{n,\cset^{\prime}_z}}.
\end{equation}

Combining these expressions with \eqref{eq:jacobian_A} and \eqref{eq:jacobian_B} (noting that we make the minor modification of setting $\cset = \rset^{\prime}$) and substituting them into \eqref{eq:cp_jacobian} we obtain
\begin{equation}
\label{eq:pjacobian_A_ele}
\begin{aligned}[b]
    [\jgumatu]_{xz}
    &= \sum_y [\jmatul]_{xy}[\jpmatu]_{yz} + \sum_y [\jmatur]_{xy}[\jpmatl]_{yz} \\
    &= - \sum_y \frac{\partial \vvec^{\vphantom{\dagger}}_{\rset_x}}{\partial \gvec^{\vphantom{\dagger}}_{n,\cset_y}}\frac{\partial \gvec^{\vphantom{\dagger}}_{n,\rset^{\prime}_y}}{\partial \uvec^{\vphantom{\dagger}}_{n,\cset^{\prime}_z}} - \sum_y \frac{\partial \vvec^{\vphantom{\dagger}}_{\rset_x}}{\partial \gvecc^{\vphantom{\dagger}}_{n,\cset_y}}\frac{\partial \gvecc^{\vphantom{\dagger}}_{n,\rset^{\prime}_y}}{\partial \uvec^{\vphantom{\dagger}}_{n,\cset^{\prime}_z}},
\end{aligned}
\end{equation}
where the number of non-zero elements entering the summation will depend on how we define $\cset^{\prime}$. The elements of $\jgumatl$ are simply given by the conjugate of the above.

As in the complex case, we want to avoid taking a large matrix inverse due to its computational complexity. Consequently, we will approximate $\ahess$ with its diagonal elements ($N \times N$ blocks). This approximation results in the following compact expression
\begin{equation}
\label{eq:jhj_entry_param}
    \begin{aligned}[b]
        [\ahess]_{n,\cset^{\prime}_i}
        &= [\jgumatuct]^{\vphantom{\dagger}}_{ix} \wmat [\jgumatu]^{\vphantom{\dagger}}_{xi} + [\jgumatlct]^{\vphantom{\dagger}}_{ix} \wmat [\jgumatl]^{\vphantom{\dagger}}_{xi} \\
        &= ([\jgumatu]^{\vphantom{\dagger}}_{xi})^{\dagger} \wmat [\jgumatu]^{\vphantom{\dagger}}_{xi} + ([\jgumatl]^{\vphantom{\dagger}}_{xi})^{\dagger} \wmat [\jgumatl]^{\vphantom{\dagger}}_{xi} \\
        &=
        2 \realop\big(([\jgumatu]^{\vphantom{\dagger}}_{xi})^{\dagger} \wmat [\jgumatu]^{\vphantom{\dagger}}_{xi}\big),
    \end{aligned}
\end{equation}
where $\realop(\cdot)$ takes the real part of its argument and arises from the property that, for a complex number $z$, $z + \bar{z} = 2\realop(z)$. We do not make further substitutions at this point in order to preserve the conciseness of this expression.

We can apply a similar procedure to obtain the following expression for the entries of $\jhwr$
\begin{equation}
\label{eq:jhwr_entry_param}
    \begin{aligned}[b]
        [\jhwr]_{n,\cset^{\prime}_i}
        &= [\jgumatuct]^{\vphantom{\dagger}}_{ix} \wmat \rvec + [\jgumatlct]^{\vphantom{\dagger}}_{ix} \wmat \bar{\rvec} \\
        &= ([\jgumatu]^{\vphantom{\dagger}}_{xi})^{\dagger} \wmat \rvec + ([\jgumatl]^{\vphantom{\dagger}}_{xi})^{\dagger} \wmat \bar{\rvec} \\
        &=
        2 \realop\big(([\jgumatu]^{\vphantom{\dagger}}_{xi})^{\dagger} \wmat \rvec\big).
    \end{aligned}
\end{equation}

\eqref{eq:jhj_entry_param} and \eqref{eq:jhwr_entry_param} can be combined to form a generic expression for the update with respect to a specific element of the parameter vector as
\begin{equation}
\label{eq:update_param}
    \begin{aligned}[b]
        \uvec_{k+1,n,\cset^{\prime}_i}  &= \uvec_{k,n,\cset^{\prime}_i} - [\ahess]^{-1}_{n,\cset^{\prime}_i} [\jhwr]^{\vphantom{-1}}_{n,\cset^{\prime}_i}.
    \end{aligned}
\end{equation}

At this juncture we are still implicitly dealing with the matrix form of the problem, as evidenced by the presence of $\cset^{\prime}$ above. However, the problem can be further simplified if we consider specific definitions of $\cset^{\prime}$.

The remainder of this section deals with the details of deriving and implementing parameterized updates for specific choices of $\cset^{\prime}$. Readers without an interest in these details are encouraged to proceed directly to $\S$~\ref{section:software}.

We stress that what follows is not an exhaustive list of possible scenarios - \eqref{eq:cp_jacobian} can be used in all cases - but simply the most useful/commonly applicable.

\subsubsection{Scenario 1}

We begin with the simplest of the possible cases - each gain element $\gvec_{n,s,a,d}$ is associated with a parameter element $\uvec_{n,s,a,d}$ and $\jmat_2$ does not introduce a summation over any index i.e. $\jpmatu$ and $\jpmatl$ are non-zero at exactly one location in each column. This scenario typically corresponds to solving for an amplitude- or phase-only gain and $\cset^{\prime}$ is defined as
\begin{equation}
    \cset^{\prime} = \sset \times \aset \times \dset = \{ (s,a,d) \mid s \in \sset, \ a \in \aset \ \mathrm{and} \ d \in \dset\}.
\end{equation}
Consequently, applying similar intuition as in the complex case, we can rewrite \eqref{eq:jhj_entry_param} as:
\begin{equation}
\label{eq:jhj_case1}
    [\ahess]_{n,s,a,d}
    =
    2\sum_{q \neq a}
    \realop
    \big(
    \emat_{n,s,aq,d}^\dagger
    \wmat_{aq,s}
    \emat_{n,s,aq,d}
    \big),
\end{equation}
where
\begin{equation}
    \emat_{n,s,aq,d} = \frac{\partial \vvec^{\vphantom{\dagger}}_{aq,s}}{\partial \gvec^{\vphantom{\dagger}}_{n,s,a,d}}
    \frac{\partial \gvec^{\vphantom{\dagger}}_{n,s,a,d}}{\partial \uvec^{\vphantom{\dagger}}_{n,s,a,d}},
\end{equation}
and we are again free to omit the $d$ from the subscripts in the case that the term in question is direction-independent. Applying a similar procedure to \eqref{eq:jhwr_entry_param} we obtain:
\begin{equation}
\label{eq:jhwr_case1}
    \begin{aligned}[b]
        [\jhwr]_{n,s,a,d} &=
        2 \sum_{q \neq a}
        \realop
        \big(
        \emat_{n,s,aq,d}^{\dagger}
        \wmat_{aq,s}
        \rvec_{aq,s}
        \big).
    \end{aligned}
\end{equation}

\eqref{eq:jhj_case1} and \eqref{eq:jhwr_case1} provide us with an easy way of implementing parameterized updates given the above definition of $\cset^{\prime}$ without requiring large matrix products or inverses.

\subsubsection{Scenario 2}

The second scenario we consider has been alluded to previously - the gains and parameters need not have the same spectral and/or temporal resolution. This is often desirable as we may benefit from the improved SNR associated with solving for the parameters over a larger interval, but still evaluate the gains at every sampled point. This is consistent with quantities such as residual/atmospheric delay which manifest as a slope in frequency. In this scenario, each $\gvec_{n,s,a,d}$ is associated with a parameter element $\uvec_{n,a,d}$ and there is a summation over sample. $\cset^{\prime}$ is defined as
\begin{equation}
    \cset^{\prime} = \aset \times \dset = \{ (a,d) \mid a \in \aset \ \mathrm{and} \ d \in \dset\}.
\end{equation}
Consequently, we can rewrite \eqref{eq:jhj_entry_param} as:
\begin{equation}
\label{eq:jhj_case2}
    [\ahess]_{n,a,d}
    =
    2 \sum_{q \neq a,s}
    \realop
    \big(
    \emat_{n,s,aq,d}^\dagger
    \wmat_{aq,s}
    \emat_{n,s,aq,d}
    \big),
\end{equation}
where
\begin{equation}
    \emat_{n,s,aq,d} = \frac{\partial \vvec^{\vphantom{\dagger}}_{aq,s}}{\partial \gvec^{\vphantom{\dagger}}_{n,s,a,d}}
    \frac{\partial \gvec^{\vphantom{\dagger}}_{n,s,a,d}}{\partial \uvec^{\vphantom{\dagger}}_{n,a,d}}
\end{equation}
and we can discriminate between the direction-dependent and direction-independent cases as before. Applying a similar procedure to \eqref{eq:jhwr_entry_param} we obtain:
\begin{equation}
\label{eq:jhwr_case2}
    \begin{aligned}[b]
        [\jhwr]_{n,a,d} &=
        2 \sum_{q \neq a,s} \realop \big(
        \emat_{n,s,aq,d}^{\dagger}
        \wmat_{aq,s}
        \rvec_{aq,s}
        \big).
    \end{aligned}
\end{equation}

\subsubsection{Scenario 3}

The third scenario we consider involves cases in which the parameter we are solving for is direction-independent e.g. pointing error, but its effect manifests as a direction-dependent gain. In this scenario, each $\gvec_{n,s,a,d}$ is associated with a parameter element $\uvec_{n,s,a}$ and $\jmat_2$ introduces a summation over direction. $\cset^{\prime}$ is defined as:
\begin{equation}
    \cset^{\prime} = \sset \times \aset = \{ (s,a) \mid s \in \sset \ \mathrm{and} \ a \in \aset\}.
\end{equation}
In this case, we can rewrite \eqref{eq:jhj_entry_param} as:
\begin{equation}
\label{eq:jhj_case3}
    [\ahess]_{n,s,a}
    =
    2 \sum_{q \neq a}
    \realop
    \big(
    \emat_{n,s,aq}^\dagger
    \wmat_{aq,s}
    \emat_{n,s,aq}
    \big),
\end{equation}
where
\begin{equation}
    \emat_{n,s,aq} = \sum_d \frac{\partial \vvec^{\vphantom{\dagger}}_{aq,s}}{\partial \gvec^{\vphantom{\dagger}}_{n,s,a,d}}
    \frac{\partial \gvec^{\vphantom{\dagger}}_{n,s,a,d}}{\partial \uvec^{\vphantom{\dagger}}_{n,a}}.
\end{equation}

Applying a similar procedure to \eqref{eq:jhwr_entry_param} we obtain:
\begin{equation}
\label{eq:jhwr_case3}
    \begin{aligned}[b]
        [\jhwr]_{n,s,a} &=
        2 \sum_{q \neq a}
        \realop
        \big(
        \emat_{n,s,aq}^\dagger
        \wmat_{aq,s}
        \rvec_{aq,s}
        \big).
    \end{aligned}
\end{equation}

It is possible to combine scenario 2 with scenario 3 by including sample index in the summation appearing in \eqref{eq:jhj_case3} and \eqref{eq:jhwr_case3}.

\section{Software}
\label{section:software}

Much of the software stack which we use has been extensively described in Paper I. However, in the interests of preserving readability and exploring our specific use case, we will briefly reintroduce these components here.

\subsection{\dask{}}
\label{subsection:dask}
\dask{} \citep{rocklin2015} is a Python package which allows Python code to be run in a distributed fashion. This is accomplished using directed acyclic graphs (or DAGs) where vertices represent tasks (operations) and edges encode the dependencies between those tasks. These DAGs can be submitted to a distributed scheduler which is responsible for mapping the vertices (tasks) of that graph onto the distributed hardware. Using this approach it is possible to scale code from running on a laptop to running on a large, possibly cloud-based, distributed system.

It is worth noting that by utilizing \dask{} (and specifically the distributed scheduler) over a technology like MPI, one forfeits fine-grained control of task placement and execution order. The benefit is increased resilience i.e. in the event that a computing resource becomes unavailable, the scheduler knows how to recover. However, experience has shown that there may be non-negligible performance degradations associated with relinquishing this control entirely. Fortunately, there exist mechanisms (so-called scheduler plugins) to manipulate how tasks are assigned to resources.

\subsection{\daskms{}}
\label{subsection:dask-ms}

\daskms{} \citep{africanus1} is a Python package that provides an interface to Measurement Sets backed by the Casacore Table Data System \citep[CTDS,][]{vandiepen2015}, the most commonly used on-disk format in radio interferometry. This package makes it possible to expose read and write operations on Measurement Sets as \dask{} arrays (chunked, lazily evaluated equivalents of \numpy{} arrays) which can be utilized during graph construction.

An additional feature of \daskms{} is experimental support for alternate on-disk representations of the Measurement Set i.e. Measurement Sets backed, not by the CTDS, but by \zarr{}\footnote{\url{https://zarr.dev/}} or Parquet\footnote{\url{https://parquet.apache.org/}}. While a detailed discussion of these formats appears in Paper I, it is worth mentioning that \zarr{} is of particular interest in the context of the calibration problem as it is a chunked format with support for massively parallel reads and writes. This feature is notably absent from the CTDS due to thread safety concerns. Additionally, \zarr{} is compatible with object stores e.g. Amazon S3, which makes it particularly suitable for cloud-based applications.

\subsection{\xarray{}}
\label{subsection:xarray}

\xarray{} \citep{hoyer2017} is a package aimed at multidimensional array labelling and manipulation in Python. Specifically, it provides intuitive mechanisms for representing arrays and collections of arrays whilst keeping track of dimension labels and, optionally, coordinates. As radio interferometric data and, indeed, gain solutions typically live on a rectilinear time-frequency grid, \xarray{} makes representing and manipulating this data simple.

\xarray{} is compatible with \dask{} arrays and is used by \daskms{} to represent the Measurement Set data concisely.

\subsection{\numba{}}
\label{subsection:numba}

It is well known that Python has become the programming language of choice in many scientific disciplines. It is not without its drawbacks as, as an interpreted language, it is not known for its speed. A particular feature of the Python Interpreter known as the Global Interpreter Lock or GIL, restricts the execution of Python code to a single thread per process. While there are situations in which Python multithreading can be useful for I/O bound operations i.e. where time is spent reading from disk, it typically provides little to no benefit in compute-bound tasks.

There are several ways in which the above problem can be ameliorated. The first is the use of \numpy{}, which drops the GIL for many of its array-based operations. However, not all problems are amenable to being expressed using arrays. The second approach is to use a combination of multiprocessing and shared memory to effectively bypass the GIL as is done in both \cubical{} and \textsc{DDFacet}, but this typically leads to more complicated, difficult to maintain code.  

Finally, it is possible to use \numba{} \citep{lam2015}: a just-in-time (JIT) compiler for a subset of Python and \numpy{}. By compiling sections of code at run-time, \numba{} can offer speed comparable with C/C++ without completely sacrificing the simplicity of Python. The code can also be compiled such that it drops the GIL, removing the limitations on multithreaded performance.  

\numba{} also allows for parallelization of code using its own internal thread pools. This lower level parallelism (in contrast to the higher level \dask{} parallelism), typically has a lower memory footprint and can be used to leverage additional CPU cores in instances where further \dask{} parallelism would lead to memory issues.

\section{Implementation}
\label{section:implementation}

Armed with the mathematics of $\S$~\ref{section:mathematics} and the packages described in $\S$~\ref{section:software}, we can finally detail the implementation of \quartical{}\footnote{\url{https://github.com/ratt-ru/QuartiCal}} \citep{quartical_ascl}, a new calibration package implemented in Python for embarrassingly parallel calibration of radio interferometer data.

\subsection{Data Ingestion}
\label{subsection:ingest}

\quartical{} makes use of \daskms{} to generate \xarray{} Dataset objects representing the data on disk. The way in which this data is represented in memory is highly configurable as \daskms{} supports both grouping and chunking.

Grouping or partitioning refers to a (typically) coarse splitting of the data based on some criteria. Common examples include partitioning by spectral window (a range of simultaneously observed frequencies), field (the target which the telescope was observing) and scan (varies from telescope to telescope, but typically a range of contiguous observation time). \daskms{} produces a single \xarray{} dataset object for each unique combination of partitioning parameters.

Chunking refers the process of representing an array, not as a single monolithic entity (a single buffer in memory), but as a rectilinear grid of chunks. These chunks may be much smaller than their parent array. The purpose of chunking is twofold: to establish work which can be done in parallel and to make larger than memory problems tractable. Chunking functionality is provided by the \dask{} array interface and exposed to users as chunked \dask{} arrays on the \xarray{} datasets provided by \daskms{}.

In \quartical{}, arrays are typically chunked in time and frequency and the extent of the chunks will usually span one or more solution intervals in each dimension. This is an important point; as the goal is to implement calibration in an embarrassingly parallel manner, we want to avoid communication between chunks. This has implications for large solution intervals, as the solution interval places a lower bound on the chunk size.

The Measurement Set introduces some additional complexity to the chunking process as it is not stored with an explicit time axis. Instead, each row of the Measurement Set is associated with a specific time (and the number of rows may be substantially greater than the number of unique times). As a result, chunking in time requires preprocessing to establish which rows are associated with specific times. This functionality is not available in \daskms{} and has to be handled in \quartical{}.

Chunking also affects performance as many chunks may be processed in parallel. This is particularly true in the distributed case where we may become I/O bound i.e. we may be unable to fully exploit our compute capacity (CPU cores) if we cannot provide them with adequate data to process. Consequently, the ability to read and write data in parallel is crucial. Unfortunately, for the reasons given in Paper I, the Measurement Set backed by the CTDS is unable to meet this requirement. \daskms{} offers an alternative by allowing conversion from a Measurement Set backed by the CTDS to a version of the same Measurement Set backed by \zarr{}. \zarr{} is a chunked, on-disk format which is particularly conducive to parallel reads and is compatible with objects stores such as Amazon S3. As \daskms{} abstracts the underlying storage layer, \quartical{} is able to use conventional and \zarr{}-backed Measurement Sets interchangeably.

One area in which \quartical{} is known to be lacking is its failure to interleave disk I/O and compute. While it is not impossible, there is no simple way to accomplish it in the current \dask{} ecosystem. Additionally, there is also an argument against this behaviour as it can introduce memory backpressure (see Paper I). In other words, it can result in flooding memory with data which cannot be processed immediately. This is particularly problematic if it results in out-of-memory errors. Further investigation of how this behaviour can be robustly incorporated in \quartical{} will be deferred to future work.

% Elected not to create this figure.
% \begin{figure}
%     \centering
%     \includegraphics[width=\columnwidth]{example-image-a}
%     \caption{Figure showing relationship between data on disk and \dask{} chunks.}
% \end{figure}

\subsection{Graph Construction}

As alluded to in $\S$~\ref{subsection:dask}, \dask{} can be used to construct task graphs which describe a computation. These task graphs can subsequently be submitted to a variety of different schedulers for execution (see $\S$~\ref{subsection:graph_execution}).

\begin{figure}
    \centering
    \includegraphics[width=0.7\columnwidth]{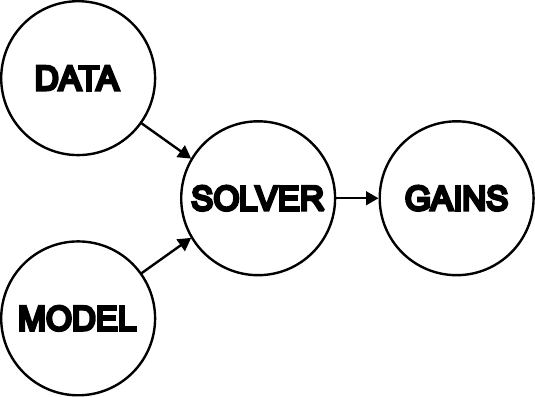}
    \caption{A simple compute graph. Circles represent nodes or tasks and the arrows show the connections between them. Conceptually, the above could represent a single chunk of the input data and input model being passed into a solver to produce a single chunk of the output gains.}
    \label{fig:basic_graph}
\end{figure}

In \quartical{}, the purpose of graph construction is to map every chunk described in $\S$~\ref{subsection:ingest} through the calibration algorithm described in $\S$~\ref{section:mathematics} in an embarrassingly parallel fashion. Practically, this entails utilizing \dask{} array functionality to map functions over array chunks. A simplified example of a task graph appears in Figure \ref{fig:basic_graph}.

In principle, this is a very simple procedure, but it is complicated by the fact that the input to the calibration algorithm is not a single array chunk. Instead, chunks of several arrays, including, but not limited to, the data, model and weights are required to perform calibration. In addition to these arrays, extensive supporting information is also necessary. This touches on one of the limitations of the \dask{} array interface: while many-to-one mappings are fully supported, i.e. passing many inputs into a function to produce a single output array, many-to-many mappings are considerably less so. This is problematic as there are many cases where returning multiple values from a function is necessary. To circumvent this problem, \quartical{} makes use of low-level \dask{} functionality to implement a class capable of expressing these many-to-many mappings during graph construction. A more in-depth discussion of this functionality is beyond the scope of this work but interested readers are encouraged to inspect the \textit{Blocker} class in the code.

As one may expect, the aforementioned many-to-many mappings rapidly complicate the graph and somewhat limit our ability to describe it in text. This complexity is also due, in part, to the fact that the graph describes the calibration process in its entirety (i.e. the graph is end-to-end). To elaborate, given the input data chunks, the graph includes all the operations necessary to map those data chunks through the calibration algorithm and ultimately write the gain solutions (in addition to any visibility products) to disk.

Implicit in this approach is a functional programming style where graph nodes encapsulate functions that transform immutable inputs to produce immutable outputs. Such functions are pure, deterministic and prevent mutation of program state. This is a double-edged sword: on the one hand it is beneficial in that it discourages the developer from building stateful applications which may be difficult to maintain and debug; on the other, it may result in spurious memory copies due to the inability of functions to modify their inputs. This, in turn, may result in a larger than necessary memory footprint.

\subsection{Solvers}
\label{subsection:solvers}

\quartical{} can solve for a number of different gain terms, both individually and in a chain\footnote{The documentation available at \url{https://quartical.readthedocs.io} includes a full list of the supported term types.}. This process is the most computationally demanding component of the package and, as such, has been implemented using \numba{}. The solution procedure for each term is fairly uniform (up to some simple modifications for parameterized terms), and follows Algorithm \ref{alg:complex}. For Jones chains including multiple terms, Algorithm \ref{alg:complex} is applied to each term individually i.e. each term is solved to convergence/the maximum number of iterations before moving onto the next. This can be repeated for multiple epochs (loops through the Jones chain). Note that Algorithm \ref{alg:complex} is a high-level abstraction: the underlying code actually evaluates $\ahess$ and $\jhwr$ in tandem without allocating the entirety of the residual vector $\breve{\rvec}$. Additionally, solution intervals are handled using mappings from the data dimensions to those of the gains as opposed to explicit broadcasting. Readers interested in these technical details are encouraged to examine the code.

\begin{algorithm}
    \caption{Approximate GN solver}\label{alg:complex}
    \begin{algorithmic}[1]
        \State \textbf{Input:} Visibilities $\dvec$, predicted visibilities $\mvec$, weights $\mathbf{w}$, initial gain estimate $\gvec_0$, tolerance $\epsilon$, maximum number of iterations $N_{\text{max}}$
        \State \textbf{Output:} Refined gain estimate $\gvec$
        \State $n \gets 0$
        \State $\gvec \gets \gvec_0$
        \While{$\|\Delta \gvec\| > \epsilon$ \textbf{and} $n < N_{\text{max}}$}
            \State Compute $\jhwr$ using \eqref{eq:jhr_entry_simplified}.
            \State Compute diagonal entries of $\ahess$ using \eqref{eq:jhj_entry_simplified}.
            \State Compute the update: $\Delta \gvec = (\ahess)^{-1}\jhwr$
            \State Update the parameter estimate: $\gvec \gets \gvec + \Delta \gvec$
            \State Increment $n$: $n \gets n + 1$
        \EndWhile
    \end{algorithmic}
\end{algorithm}

On the topic of avoiding allocation, \quartical{} excels relative to its predecessor \cubical{}: its solvers use substantially less memory (see $\S$~\ref{section:results}). This is accomplished by avoiding all visibility-sized intermediaries i.e. we never attempt to keep large arrays/array chunks in memory with the obvious exception of the actual visibility inputs and weights. This does come at a computational cost as we need to repeatedly evaluate potentially lengthy Jones chains (consisting of $\mathcal{O}(n^3)$ matrix multiplications). However, our experience has shown that excessive memory footprint ultimately results in slower implementations as it may become difficult to leverage computational resources (CPU cores) due to the finite amount of work that can be done in parallel. Additionally, minimizing memory footprint makes it possible to deal with larger problems on less powerful hardware. This is beneficial when attempting to use cloud-based resources, where more powerful hardware has a commensurate cost.

In order to account for some of the problems alluded to in $\S$~\ref{subsection:ingest}, specifically the case where large solution intervals limit the use of \dask{} threads (due to their associated memory footprint), \quartical{}'s solvers make use of \numba{}'s parallel range (prange) functionality. This is used in conjunction with \dask{} to implement a form of nested parallelism where each \dask{} thread may in turn spawn several \numba{} threads. \dask{} parallelism tends to be more efficient as all operations (which do not require the GIL) may be done in parallel, whereas the \numba{} parallelism only applies to the code in which it is implemented. That said, prange functionality has been included in all the most computationally demanding components of the solvers (specifically the computation of $\ahess$ and $\jhwr$). This allows \quartical{} to use multiple CPU cores even in cases where only a single chunk of data fits in memory. Crucially, the \numba{}-based parallelism in \quartical{} incurs almost no additional memory footprint.

\quartical{}'s solvers also include a handful of features not mentioned in Algorithm \ref{alg:complex} for the sake of simplicity. These include:
\begin{itemize}
    \item Gain flagging: \quartical{} attempts to identify and flag any gain solutions which appear to be diverging i.e. for which the magnitude of the gain updates does not tend to zero.
    \item Initial estimates: Certain parameterized terms e.g. delay, can be estimated using alternative techniques. In the case of the delay, this is accomplished by finding the peaks in the Fourier transform of the baselines including the reference antenna \citep{cotton1995}.
    \item Robust reweighting: \quartical{} optionally implements robust reweighting based on \citet{sob2020} between solver epochs. This can be used to reduce overfitting of RFI and unmodelled emission.
\end{itemize}

\subsection{Outputs}
As calibration software, \quartical{}'s main aim is the production of gain solutions. There is no universally accepted format for these solutions, and each calibration package typically implements its own. The same is true of \quartical{}, as we have resolved to store our gain solutions in \xarray{} datasets backed by \zarr{} on disk. The reason for this choice was mentioned in $\S$~\ref{subsection:xarray} and follows the same reasoning behind its adoption in \daskms{}: the ability to associate the dimensions of multidimensional arrays with coordinates in a self-describing dataset is incredibly powerful. Additionally, \xarray{} itself also provides many sophisticated routines for manipulating these dataset objects such as selection and concatenation by coordinate.

\quartical{}'s gain solutions are represented by 5-dimensional arrays and \xarray{} is used to associate each dimension with the relevant coordinates (see Figure \ref{fig:xarray_dataset}). These coordinates can have any data type e.g. strings as in the case for antenna names or correlation labels, or floats containing the true physical values as in the case for the times and frequencies. In addition to the data (gains and gain flags in the example) and coordinates, a variety of additional information/metadata can be included as attributes on the \xarray{} dataset. In Figure \ref{fig:xarray_dataset}, these include the partitioning information i.e. the field, data descriptor ID and scan number the gain array is associated with as well as some information pertaining to the gain type.

\begin{figure*}
\begin{verbatim}
    <xarray.Dataset> Size: 13MB
    Dimensions:      (gain_time: 115, gain_freq: 64, antenna: 28, direction: 1,
                    correlation: 4)
    Coordinates:
    * antenna      (antenna) <U2 224B `1' `2' `3' `4' `5' ... `25' `26' `27' `28'
    * correlation  (correlation) <U2 32B `RR' `RL' `LR' `LL'
    * direction    (direction) int32 4B 0
    * gain_freq    (gain_freq) float64 512B 1.266e+09 1.27e+09 ... 1.518e+09
    * gain_time    (gain_time) float64 920B 4.884e+09 4.884e+09 ... 4.884e+09
    Data variables:
        gains        (gain_time, gain_freq, antenna, direction, correlation) complex128 13MB ...
        gain_flags   (gain_time, gain_freq, antenna, direction) int8 206kB 1 1 ... 0
    Attributes:
        DATA_DESC_ID:  0
        FIELD_ID:      0
        FIELD_NAME:    J0542+4951
        NAME:          G
        SCAN_NUMBER:   3
        TYPE:          complex
\end{verbatim}
\caption{The \xarray{} representation of \quartical{}'s gain format. Note that the above has been slightly simplified for clarity.}
\label{fig:xarray_dataset}
\end{figure*}

As \zarr{} is one of several on-disk formats natively supported by \xarray{}, the dataset format also makes loading gains simple. \quartical{} uses this functionality to allow users to load and apply (potentially on the fly upsampled/downsampled) gain solutions from previous runs and optionally interpolate solutions from one field to another. This, in turn, allows \quartical{} to handle 1GC or transfer calibration. We note in passing that we will not demonstrate this functionality in this paper, but that it will be used in the following paper in this series. Additionally, improving interpolation and its close neighbour, solution smoothing, will be pursued in a future paper.

It is our hope that this format becomes more widely adopted in future, allowing for greater interoperability between calibration packages (and imagers). This is already in the process of being realized as the next paper in this series details \pfb{} \citep{africanus3}, an imaging framework which understands \quartical{}'s gain format and can apply the gains on the fly during imaging. This makes it possible to avoid writing out additional visibility-sized outputs to the Measurement Set such as corrected data (data with the inverse of the gain solutions applied to it).

On the subject of visibility-sized outputs, \quartical{} does include the functionality to produce and write many of these products back to arbitrarily named Measurement Set columns. This includes corrected data, corrected residuals and corrected weights. While this behaviour is entirely optional (and disabled by default), it has been shown to be very useful in the past \citep[see][]{smirnov2024}, particularly when attempting to peel (subtract) problematic sources affected by direction-dependent effects.

\subsection{Graph Execution}
\label{subsection:graph_execution}

\dask{} supports several schedulers for the purpose of graph execution. These schedulers are responsible for mapping graph nodes or tasks to specific hardware and determining the order in which tasks are processed. They are an integral part of the \dask{} ecosystem and the performance of an application is inextricably tied to their behaviour. The \dask{} schedulers supported by \quartical{} are as follows:
\begin{itemize}
    \item Single-threaded scheduler: This scheduler is primarily useful for debugging as it executes the task graph serially using a single thread.
    \item Threaded scheduler: This scheduler makes use of multiple \dask{} threads on a single compute node to execute task graphs in parallel. The order in which tasks are computed are determined by \dask{}'s internal ordering routines which use static graph analysis to assign every graph node an integer priority. The threaded scheduler typically performs well and with limited memory overhead but is limited to a single compute node.
    \item Distributed scheduler: This is the most sophisticated of \dask{}'s schedulers and allows task graphs to be executed on distributed hardware using a combination of \dask{} workers and threads. The order in which tasks get executed is non-deterministic as it is dependent on the current cluster state. We will return to the repercussions of this point shortly.
\end{itemize}

Due to the embarrassingly parallel structure of \quartical{}'s \dask{} graph (characterized by many parallel, independent sub-graphs), it is possible for many tasks to be processed simultaneously when using \dask{}'s distributed or threaded schedulers. The number of tasks which will be processed in parallel is determined by the number of threads when using the multithreaded scheduler, or the product of \dask{} workers and \dask{} threads in the distributed case. A \dask{} worker is a Python process which may in turn run one or more \dask{} threads. A conventional pattern when using the distributed scheduler is to associate one \dask{} worker with each node in a cluster and, optionally, several \dask{} threads with each worker (resources permitting). The distributed case is of particular interest as, in the absence of other constraints, we can imagine solving very large problems simply by scaling horizontally (adding additional \dask{} workers).

One limitation of \dask{}'s parallelism is that it cannot circumvent the GIL i.e. pure Python code will not typically parallelize well using \dask{} threads. Fortunately, code written using either \numpy{} or \numba{} will typically drop the GIL, and this accounts for the majority of \quartical{}'s code base. Additionally, as mentioned in $\S$~\ref{subsection:solvers}, parallelism using \dask{} workers and threads has a larger memory footprint as increased \dask{}-based parallelism means more tasks will be being processed simultaneously. On hardware with memory constraints, this can limit the amount of \dask{}-based parallelism which can be used without resulting in the use of swap space or \dask{}'s spill-to-disk functionality. Both of these are typically the death-knell of an application, and \quartical{} routinely disables the latter.

Returning to the last bullet point above, while it is convenient to allow \dask{}'s distributed scheduler to handle task placement when using a cluster, in reality this may not be optimal. This stems from the fact that \dask{} doesn't attempt to reason about data locality, particularly in the long term. Consequently, the distributed scheduler will, by default, be excessively greedy in the way it evaluates tasks and will readily move data between cluster nodes even when doing so will necessitate holding additional replicas (copies) of the data in memory. This can lead to unstable and unpredictable memory behaviour which will vary from run to run. This may not be a problem on systems with sufficient overhead to accommodate this variability, but may cause out-of-memory errors on more constrained systems.

The distributed scheduler is constantly being improved by the \dask{} developers and \quartical{} performs far better using the default configuration of the distributed scheduler than it used to. However, there remain situations where is it still beneficial to wrest some control back from the scheduler. As described in Paper I, it is possible to write scheduler plugins which manipulate the way in which the distributed scheduler assigns tasks to nodes. Two such plugins have been developed for \quartical{}:
\begin{enumerate}
    \item AutoRestrictor: This plugin leverages a priori knowledge of the graph structure to forcibly assign data partitions (see $\S$~\ref{subsection:ingest}) to specific cluster workers. This almost completely eliminates unnecessary data transfers and can have a large impact on performance. The downside of this approach is that it reduces resilience (the ability of the cluster to recover from errors) due to the strict mapping from task to worker node, and places an upper bound on the amount of \dask{}-based parallelism. This bound is determined by the number of data partitions. Note that this plugin is available and enabled by default in the release version of \quartical{}.
    \item SolverRestrictor: This experimental scheduler plugin is less strict than its sibling, and only enforces strict task placement for solver tasks.  The placement of the solver tasks is then used to determine the best workers on which to run both the solver's dependents (tasks which depend on the solver) and dependencies (tasks on which the solver depends). This results in fewer large data transfers between nodes and typically ensures stable memory footprint across the cluster. Additionally, this raises the upper bound on the possible degree of \dask{}-based parallelism by allowing each chunk (as opposed to each partition) to be mapped to a different cluster worker. This plugin is still in development and has not yet been included in a release version.
\end{enumerate}

\subsection{CLI and \stimela{} Integration}

\quartical{} is a command line based application. As such, it sports a Command Line Interface (CLI) which allows users to configure its various options. However, due to \quartical{}'s flexibility, and the large number of options it provides, the standard Python CLI tools are ill-suited for this purpose. Specifically, \quartical{} needs to combine user configuration from configuration files (in .yaml format) with command line arguments while simultaneously supporting dynamically generated configuration fields which are required to enable the specification of arbitrary Jones chains.

Whilst this was originally accomplished using \textsc{OmegaConf}\footnote{\url{https://github.com/omry/omegaconf}}, this functionality has since been integrated into \stimela{}, a generic pipelining framework targeted at radio astronomy applications and the topic of Paper IV in this series \citep{africanus4}. This integration allows \stimela{} to ``speak'' to \quartical{} directly, obviating the need for files which map \quartical{} parameters into something understandable by \stimela{}. \quartical{} also benefits from not needing to maintain all the complicated code required to provide its CLI. Further elucidation of \stimela{} and its features is left to Paper IV.

\subsection{Additional Features}

Finally, \quartical{} includes some additional features which are worth mentioning. The first is flagging functionality which makes use of the Median Absolute Deviation (MAD) to identify outliers in the residual visibilities after calibration. This approach is motivated by the assumption that, for well-calibrated data in the presence of a complete or nearly complete model, the residuals should be Gaussian. Any outliers are then likely to be RFI. This flagging can be applied either globally i.e. by examining the statistics of all the baselines, or in a per-baseline fashion. This has been found to be useful in practice but requires a degree of caution as using the MAD flagger with an incomplete model will tend to flag unmodelled real emission.

A second useful feature is support for predicting component models on the fly using the DFT as implemented in \textsc{codex-africanus} \citep{africanus1}. At the time of writing, \quartical{} supports Tigger sky models\footnote{\url{https://github.com/ratt-ru/tigger-lsm}} but this functionality could easily be expanded to include alternative formats. The ability to predict visibilities is crucial for direction-dependent calibration, and support for on the fly degridding (generating visibilities from model images) is currently in development. This functionality will be included in a future release.

\section{Results}
\label{section:results}
Having explained the mathematics behind \quartical{} as well as the most relevant details of its implementation, all that remains is to demonstrate that it produces sensible gain solutions and quantify its performance in a variety of different situations.

\subsection{Qualitative Results}
\label{subsection:qualitative_results}
It is pointless discussing \quartical{}'s performance characteristics before demonstrating that it is capable of calibrating radio interferometric data. The test data in this instance was originally described as the observation labelled L1 in \citet{smirnov2024}. For reference, this was an L-band, 4096 channel, 8 second integration MeerKAT observation of the field surrounding PSR J2009-2026, an unusual pulsar-like object detected serendipitously during an observation of the Great Conjunction of Jupiter and Saturn in 2020. \quartical{} was actually used during the data reduction presented in \citet{smirnov2024}, but we reproduce (and slightly improve upon) those results here.

The data was preprocessed using the \textsc{CARACal} \citep{jozsa2020} pipeline. This preprocessing included flagging, the application of 1GC solutions (derived using CASA) and averaging from the original 4096 channels down to 1024. The resulting Measurement Set contains a little over 230GB of visibility data in its DATA column. \stimela{} was used to orchestrate and run the relevant imaging and calibration steps on the preprocessed data.

\begin{figure*}
    \centering
    \begin{subfigure}[t]{\columnwidth}
        \centering
        \includegraphics[width=\columnwidth]{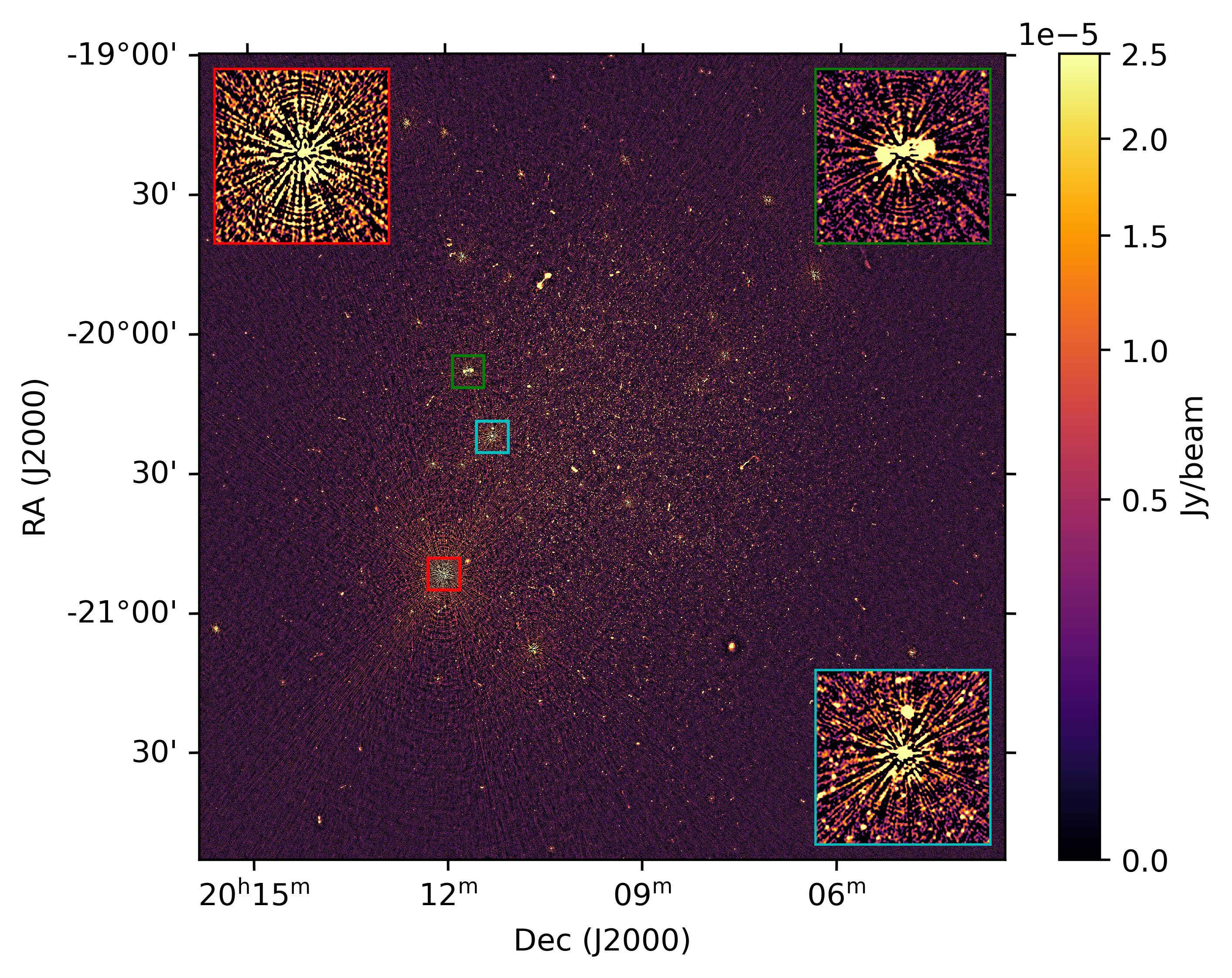}
    \end{subfigure}%
    ~ 
    \begin{subfigure}[t]{\columnwidth}
        \centering
        \includegraphics[width=\columnwidth]{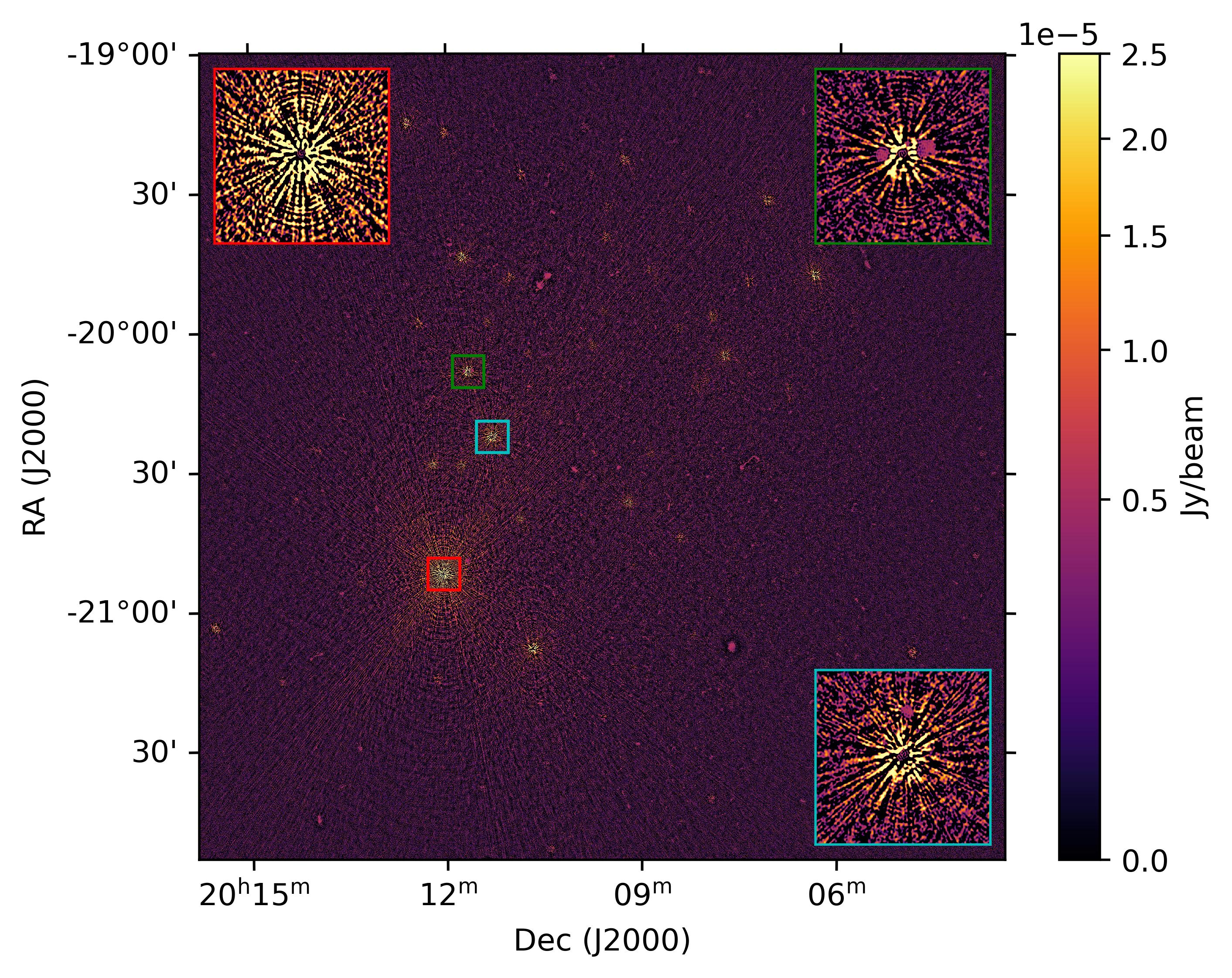}
    \end{subfigure}
    \centering
    \begin{subfigure}[t]{\columnwidth}
        \centering
        \includegraphics[width=\columnwidth]{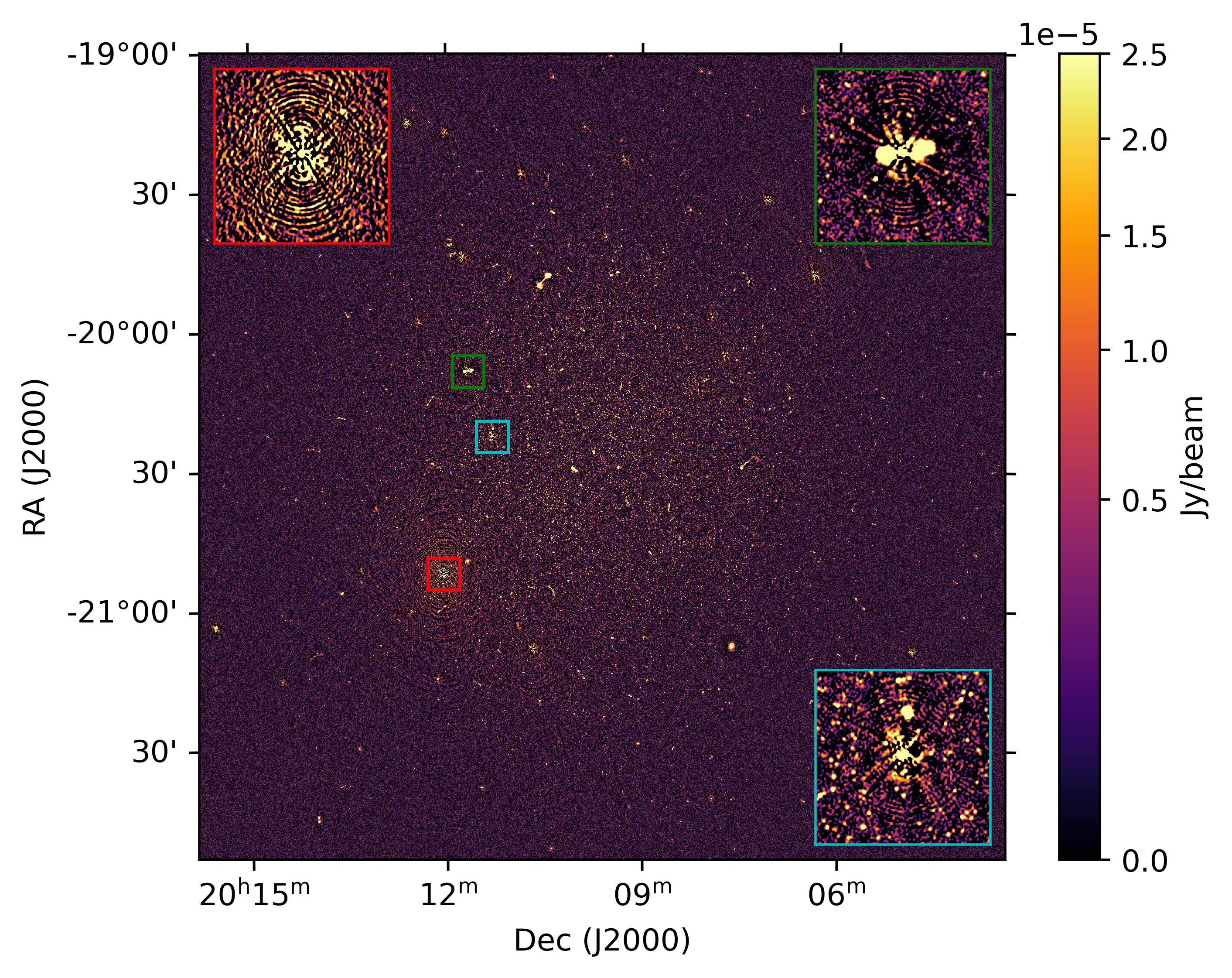}
    \end{subfigure}%
    ~ 
    \begin{subfigure}[t]{\columnwidth}
        \centering
        \includegraphics[width=\columnwidth]{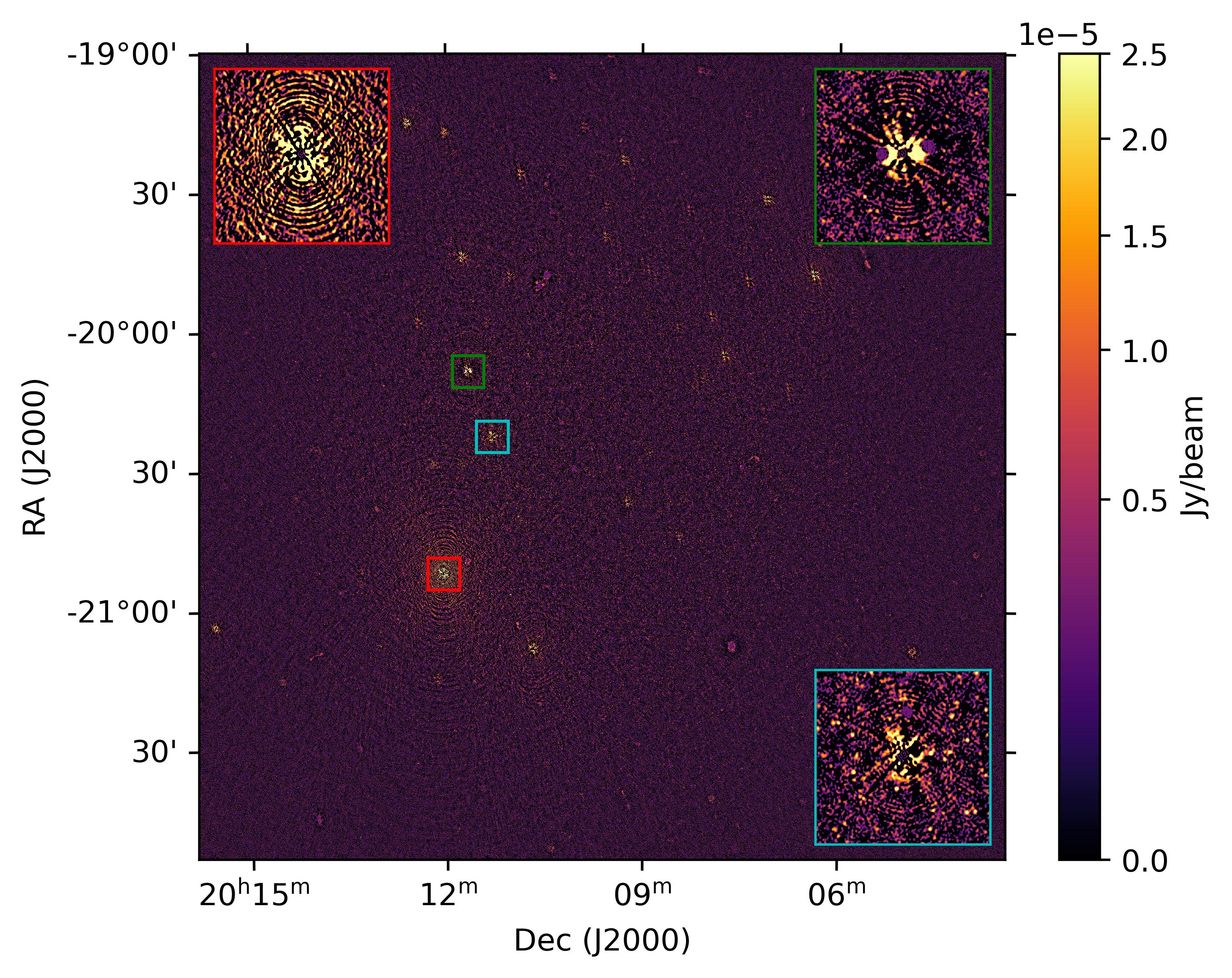}
    \end{subfigure}
    \centering
    \begin{subfigure}[t]{\columnwidth}
        \centering
        \includegraphics[width=\columnwidth]{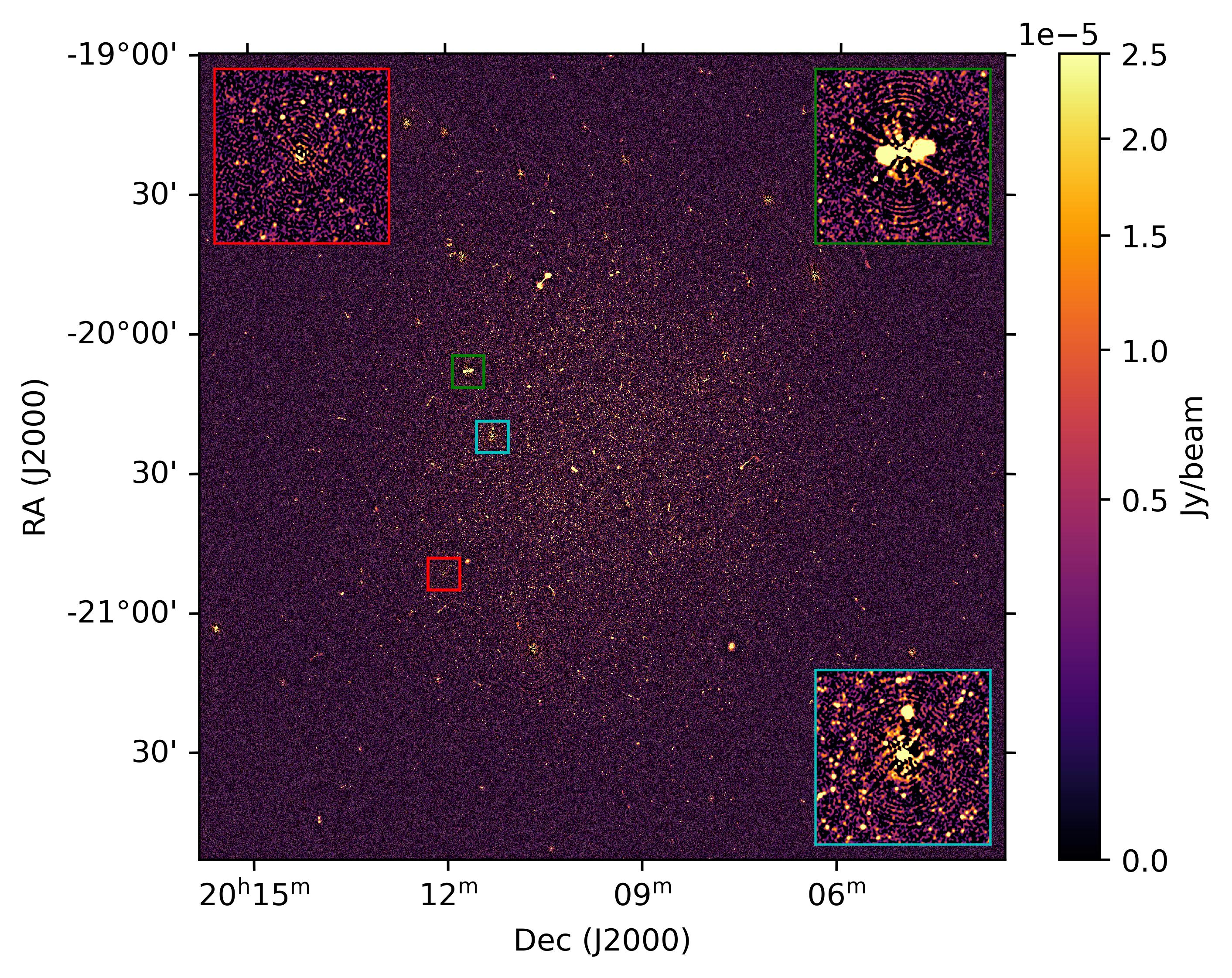}
    \end{subfigure}%
    ~ 
    \begin{subfigure}[t]{\columnwidth}
        \centering
        \includegraphics[width=\columnwidth]{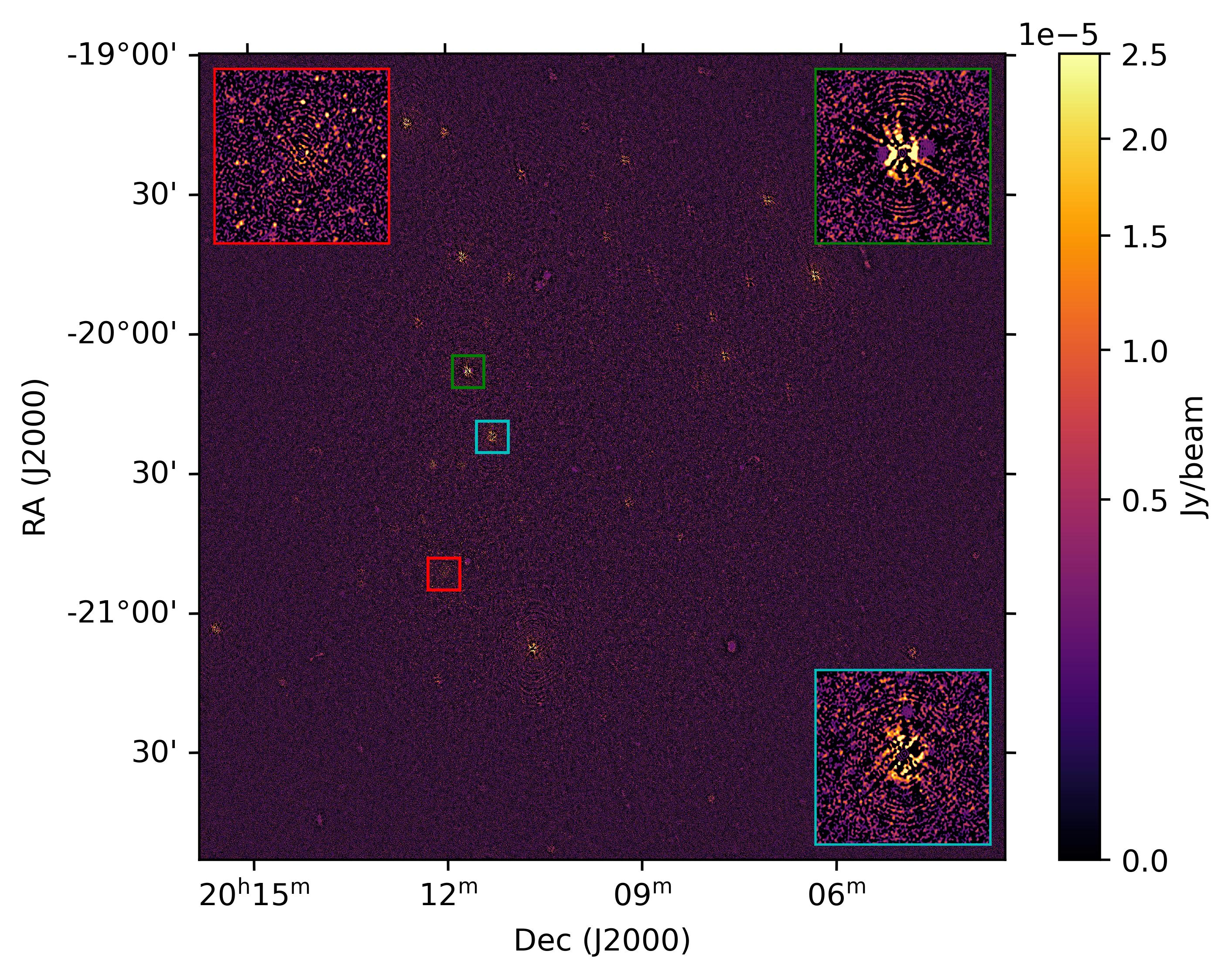}
    \end{subfigure}
    \caption{Images of the field surrounding PSR J2009-2026. The zoomed regions are colour-coded and correspond to interesting sources. \textit{Top row:} Restored (\textit{left}) and residual (\textit{right}) images prior to self-calibration. \textit{Middle row:} Restored (\textit{left}) and residual (\textit{right}) images after self-calibration with a residual delay and phase term (\textbf{K}). \textit{Bottom row:} Restored (\textit{left}) and residual (\textit{right}) images after self-calibration with \textbf{K} and a direction-dependent, diagonal, complex-valued term (\textbf{dE}). Note how the red-bordered source has been subtracted.}
    \label{fig:parrot_images}
\end{figure*}

The first outputs of this pipeline are the restored and residual images (with zoomed cutouts of interesting sources) prior to self-calibration which appear in the first row of Figure \ref{fig:parrot_images}. These 13000-by-13000 pixel images were produced using the single-scale CLEAN algorithm \citep{hogbom1974} implemented in \textsc{wsclean} \citep{offringa2014} with a cell size of 0.8 arcseconds. The component model produced by \textsc{wsclean} when making these initial images was used to populate the MODEL\_DATA column with model visibilities for use during the self-calibration process.

The next step in this process was using \quartical{} to derive gain solutions. Specifically, for the direction-independent component of the process, it was used to solve for a residual delay and phase term (denoted by \textbf{K}) over all the channels for each integration. This resulted in a modest improvement in the images (second row of Figure \ref{fig:parrot_images}), particularly around the brighter sources. The images derived after this first application of \quartical{} were used to refine the model of the field for use in subsequent rounds of calibration. However, direction-independent self-calibration can only go so far and a bright, off-axis source (red border zoom in Figure \ref{fig:parrot_images}) was found to cause significant artefacts in the resulting images. This is likely due to the effect of the primary beam, which was not modelled during this experiment. 

In order to suppress these artefacts, \quartical{} was used to peel (subtract) the offending source. This was done using a Jones chain consisting of our previously defined \textbf{K} term and a direction-dependent, diagonal, complex-valued term (denoted by \textbf{dE}), solved over 128 channels and 16 integrations. A diagonal term was used as no attempt was made to correct for beam-related leakage due to the lack of a polarized model for the field. \textsc{crystalball} (see Paper I) was used to predict the model visibilities associated with this source into a Measurement Set column for use by \quartical{}.

Images of the field after direction-dependent calibration appear in the final row of Figure \ref{fig:parrot_images}. The source was successfully, if slightly imperfectly, peeled. This imperfection, characterised by some low-level emission left in the image, can be attributed to the use of solution intervals i.e. a solution interval which contains adequate signal-to-noise to constrain a solution may be too large to fully capture the intrinsic variability of the gains. Improving this behaviour is ongoing work. The structures visible in the residuals around the green and cyan bordered regions are due to uncalibrated direction-dependent errors and flux which falls outside the mask used during deconvolution. However, their extent is limited, and further calibration was deemed unnecessary to do science at the centre of the field. In principle, these sources could also be peeled and the images further improved.

It is important to mention that we do not include a quantitative comparison between \quartical{} and its competitors for the simple reason that those competitors do not offer the same term types or the ability to chain them together in the way that \quartical{} does. As such, performing a comparison that would be fair to both pieces of software is impossible. Instead, we note that \cubical{}, the package from which \quartical{} has evolved, has been used successfully in real applications for several years \citep[e.g.][]{parekh2021,klutse2024}.

\subsection{Profiling: \quartical{} vs \cubical{}}
\label{subsection:qcvscc}

In order to demonstrate the performance characteristics of \quartical{} relative to its predecessor \cubical{}, both packages were used to process the same data as described in $\S$~\ref{subsection:qualitative_results}. While a completely fair comparison is impossible due to the differences between the packages, the results should be sufficiently conclusive to allay any such concerns.

Two experiments were devised, replicating the two main calibration steps of $\S$~\ref{subsection:qualitative_results}. The first experiment entailed solving for a residual delay and phase term (\textbf{K}) over all channels for each integration, while the second consisted of a Jones chain of the aforementioned \textbf{K} term and a direction-dependent, diagonal, complex-valued term (\textbf{dE}) solved over 128 channels and 16 integrations. Readers familiar with \cubical{}'s limitations will likely have noticed that the second case cannot be accomplished with a single run of \cubical{} as it does not support parameterized terms in a chain. Consequently, in order to keep the experiments approximately consistent, \cubical{} was used to solve for a second diagonal, complex-valued term instead of a residual delay and phase term. While this does not produce the same gain solutions as \quartical{}, it should be roughly as challenging from a numerical perspective. Additionally, \quartical{} was configured to select only the parallel hand visibilities during calibration. This functionality does not work as expected in \cubical{}.

\begin{figure*}
    \centering
    \begin{subfigure}[t]{\columnwidth}
        \centering
        \includegraphics[width=\columnwidth]{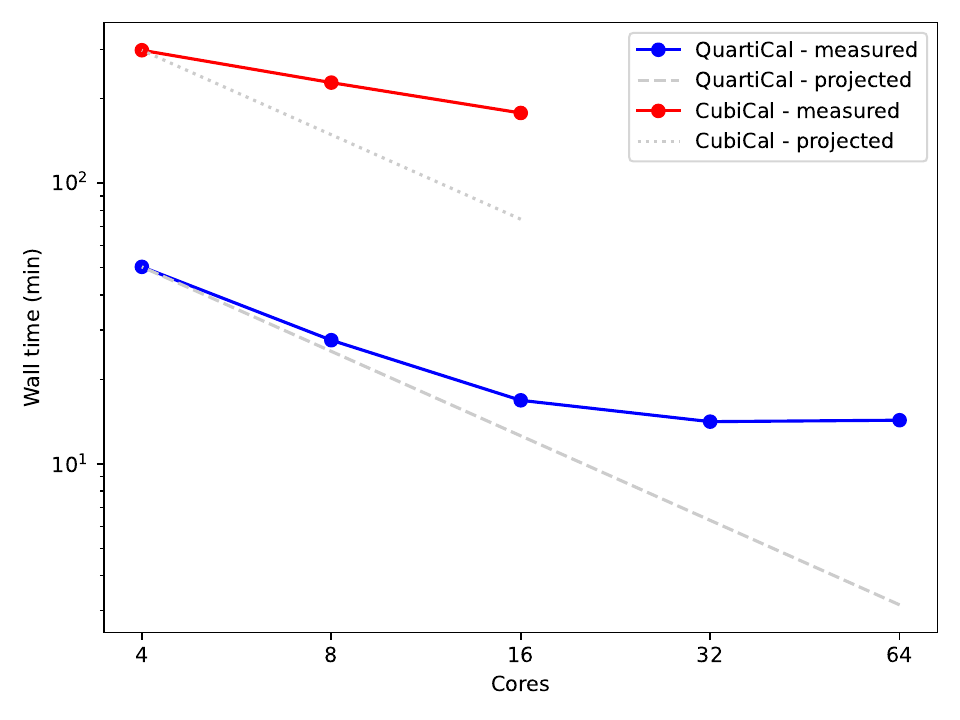}
    \end{subfigure}%
    ~ 
    \begin{subfigure}[t]{\columnwidth}
        \centering
        \includegraphics[width=\columnwidth]{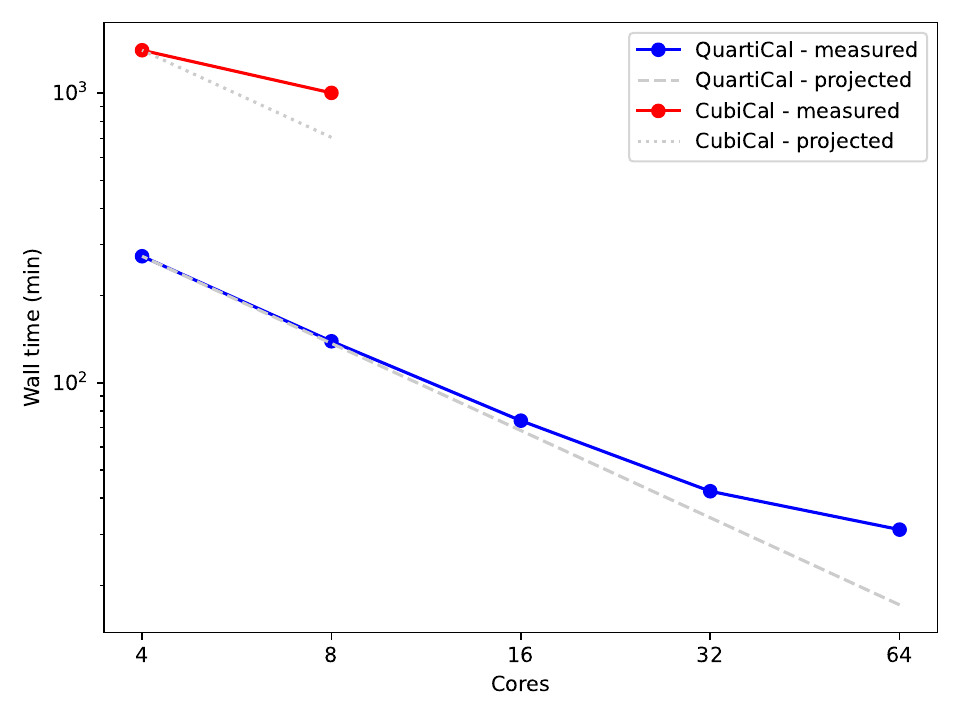}
    \end{subfigure}
    \begin{subfigure}[t]{\columnwidth}
        \centering\captionsetup{width=.8\linewidth}
        \includegraphics[width=\columnwidth]{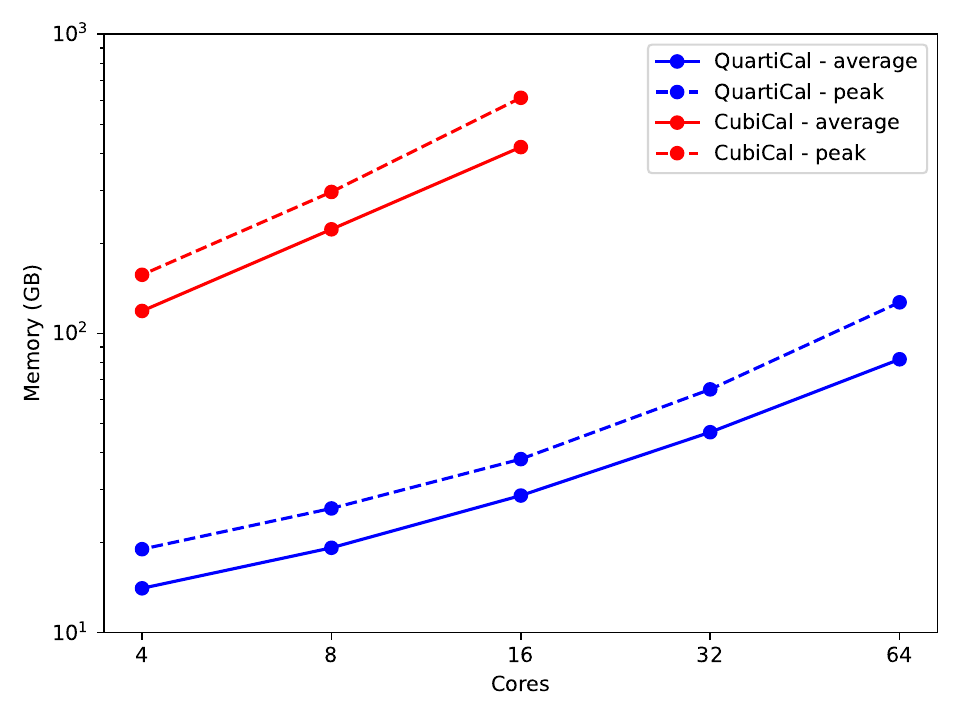}
        \caption{Solving for residual delays and phase offsets (\textbf{K}).}
    \end{subfigure}
    ~ 
    \begin{subfigure}[t]{\columnwidth}
        \centering\captionsetup{width=.8\linewidth}
        \includegraphics[width=\columnwidth]{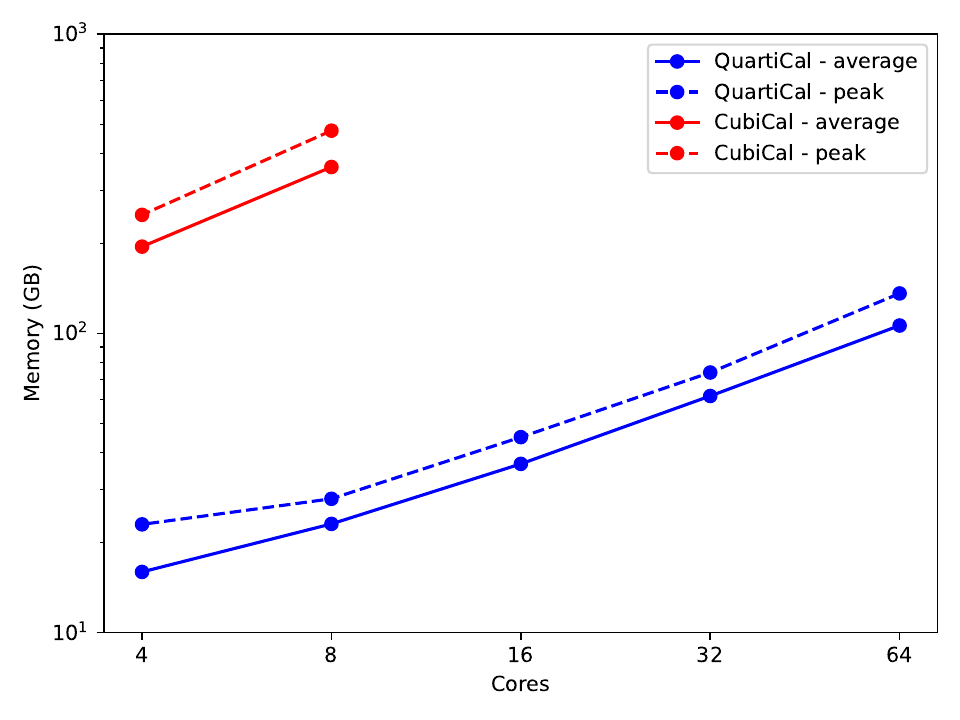}
        \caption{Solving for residual delays and phase offsets (\textbf{K}) and a direction-dependent, diagonal, complex-valued gain (\textbf{dE}).}
    \end{subfigure}
    \caption{Benchmarking results comparing \quartical{} and \cubical{} when solving for \textbf{K} (\textit{left}) and \textbf{KdE} (\textit{right}). \textit{Top row:} Wall time in minutes as a function of allocated CPU cores. The projected curve is generated by presuming linear scaling and extrapolating from the measured value for the 4 core case. \textit{Bottom row:} Memory footprint in gigabytes as a function of allocated CPU cores.}
    \label{fig:qc_vs_cc}
\end{figure*}

These experiments were conducted on a single large node containing an AMD EPYC 7773X 64-Core Processor (with simultaneous multithreading disabled), 1TB of RAM and an SSD-backed file system. For each package, the number of cores in use (by threads in \quartical{} and processes in \cubical{}) was scaled from 4 to 64 in powers of two. The \quartical{} results were generated using \dask{}'s threaded scheduler (see $\S$~\ref{subsection:graph_execution}). The wall time as well as the peak and average memory usage of the two packages were measured as a function of the number cores in use. This was done using the run statistics generated by \stimela{}, which include (among other things) these exact values. These statistics are gathered at 1-second intervals and provide a convenient way of comparing the packages. Attempts were made to quantify the relative CPU usage of each package, but the results were difficult to interpret due to the use of multiprocessing in \cubical{}. Plots of the wall time and memory usage for both experiments appear in Figure \ref{fig:qc_vs_cc}.

\quartical{} outperforms \cubical{} in every way, but most notably with respect to memory footprint. This (approximately) order of magnitude improvement can be attributed to \quartical{} avoiding large intermediary results and its utilization of mappings to avoid having to coerce the data onto an explicit rectilinear time-frequency grid. \quartical{}'s memory usage appears to scale approximately linearly as a function of the allocated cores. Naturally, there are some overheads, and the fact that the peak memory usage differs from the average memory usage is largely a result of the functional programming paradigm, i.e. there are cases where additional arrays are allocated in order to prevent the mutation of inputs. We are still considering approaches which will allow further reduction in the peak memory footprint.

It is noteworthy that \quartical{} outperforms \cubical{} in terms of wall time as, while it is not clear in the plots, in order for the \cubical{} experiments to complete in a reasonable amount of time, we were forced to reduce the total number of iterations, i.e. \cubical{} had fewer solution epochs. As such, \quartical{} did more work in less time than its competitor. Additionally, we draw the reader's attention to the fact that there are far fewer data points for \cubical{} than for \quartical{}. This is due to the fact that \cubical{}'s substantially larger memory footprint results in it using swap space for these problem sizes. Consequently, the processing time deteriorated to the point that it was impossible to allow those experiments to finish without crippling the node for days to weeks. This should drive home the point that \quartical{} succeeds in maintaining a reasonable memory footprint while simultaneously making a high degree of parallelism possible.

For high core counts, the wall time plots indicate that \quartical{} does not scale linearly. This is disappointing but there are a couple of potential causes for this behaviour. The first is that, in the case where the Jones chain only includes \textbf{K}, there is substantial I/O overhead relative to the amount of compute. Consequently, \quartical{} struggles to saturate the node with adequate work and performance decays. This is suggested by the fact that the wall time does not decrease from the 32 core result to the 64 core result. This is also consistent with the behaviour in the case of the \textbf{KdE} chain, where there is substantially more compute required i.e. the I/O is a lesser fraction of the overall run time and the deviation from linear scaling is less pronounced.

The second potential cause stems from the fact that there is only a finite amount of data to process. As \quartical{} processes chunks of data in parallel, if there are relatively few chunks of data to process per core, our performance results will be skewed by the behaviour of the worst performing chunks. This is in addition to the fact that the number of chunks may not be perfectly divisible by the number of cores in use i.e. some cores will end up doing more work. For this particular Measurement Set, the data is represented as 239 chunks. This means that in the 64 core case, each core will process approximately 3.7 chunks of data. Another way of thinking about this is that 70\% of the cores will process four chunks of data while 30\% of the cores will only process three. As the wall time is simply a measure of the total time taken to finish calibrating the data, at high core counts we enter a regime in which our wall time measurements may be dominated by the chunks which are the slowest to converge and the cores which have to do the most work. It is worth stressing that these experiments involve real data and an iterative algorithm. Consequently, the processing time of each chunk may vary significantly.

\subsection{Profiling: CTDS-backed Measurement Set vs \zarr{}-backed Measurement Set}
\begin{figure*}
    \centering
    \begin{subfigure}[t]{\columnwidth}
        \centering
        \includegraphics[width=\columnwidth]{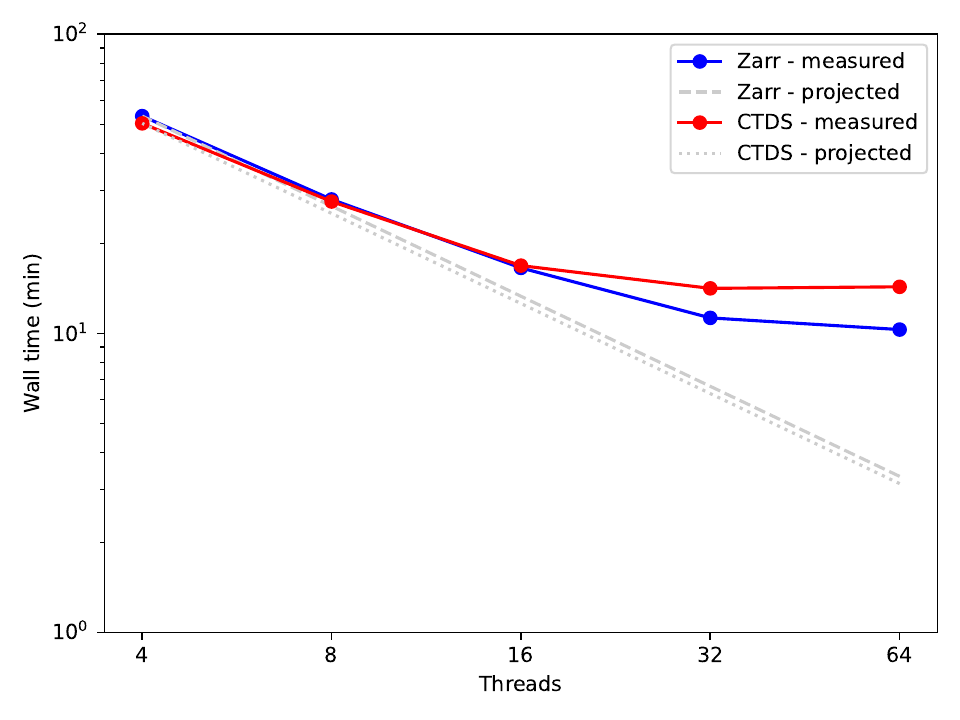}
    \end{subfigure}%
    ~ 
    \begin{subfigure}[t]{\columnwidth}
        \centering
        \includegraphics[width=\columnwidth]{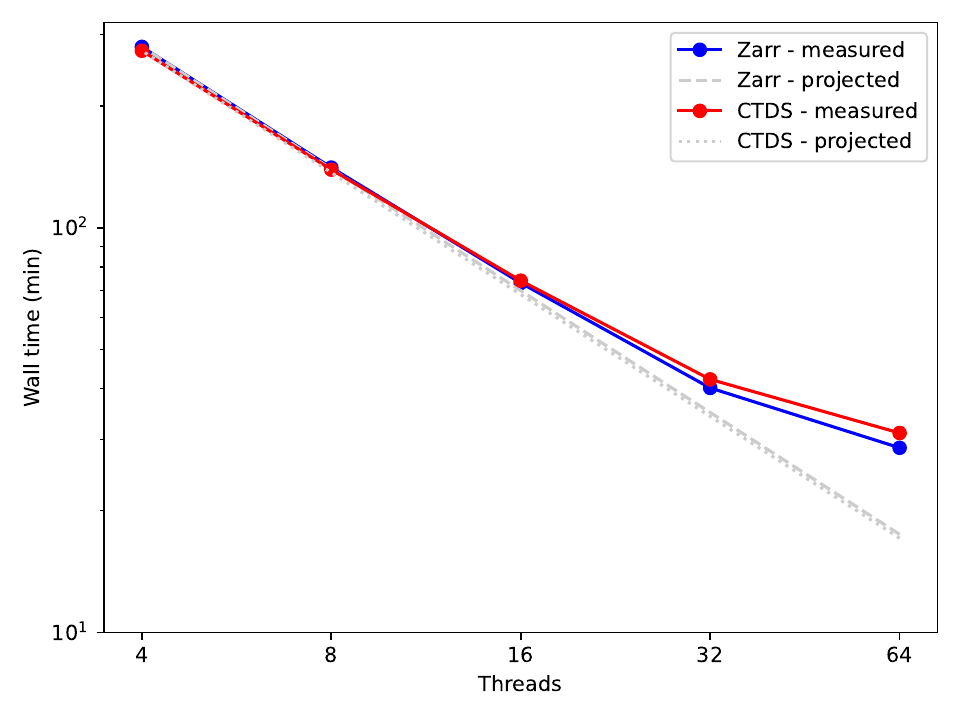}
    \end{subfigure}
    \centering
    \begin{subfigure}[t]{\columnwidth}
        \centering
        \includegraphics[width=\columnwidth]{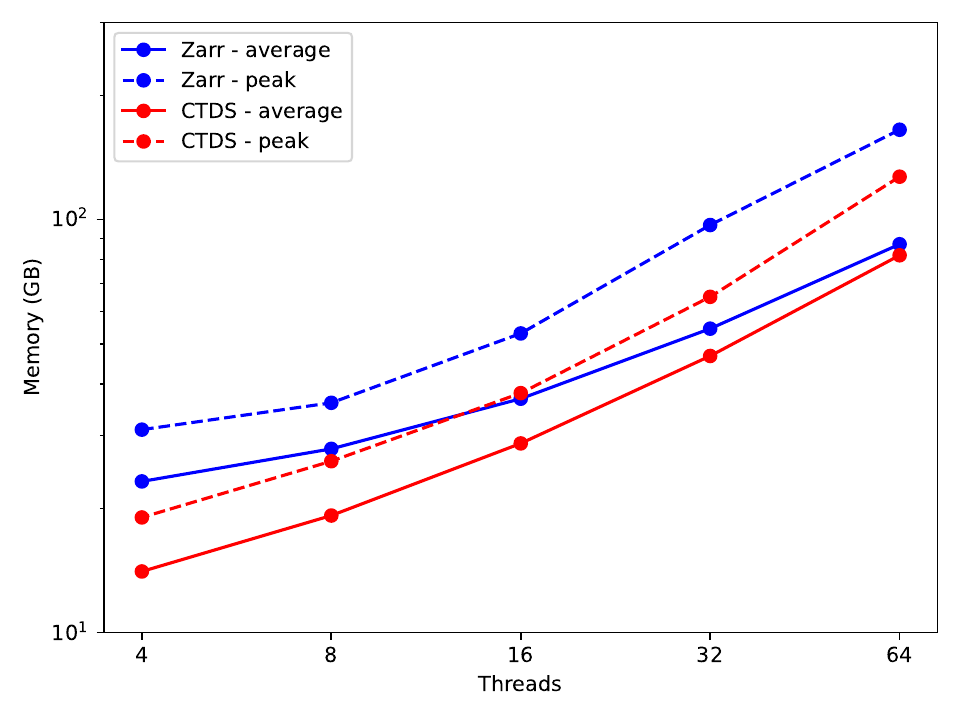}
    \end{subfigure}%
    ~ 
    \begin{subfigure}[t]{\columnwidth}
        \centering
        \includegraphics[width=\columnwidth]{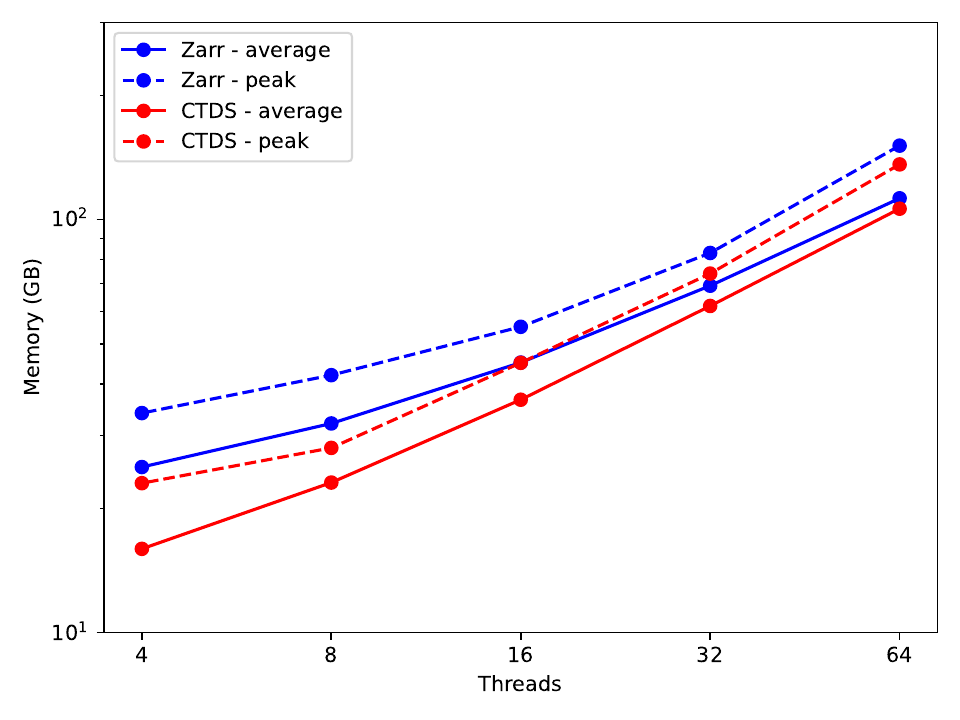}
    \end{subfigure}
    \centering
    \begin{subfigure}[t]{\columnwidth}
        \centering\captionsetup{width=.8\linewidth}
        \includegraphics[width=\columnwidth]{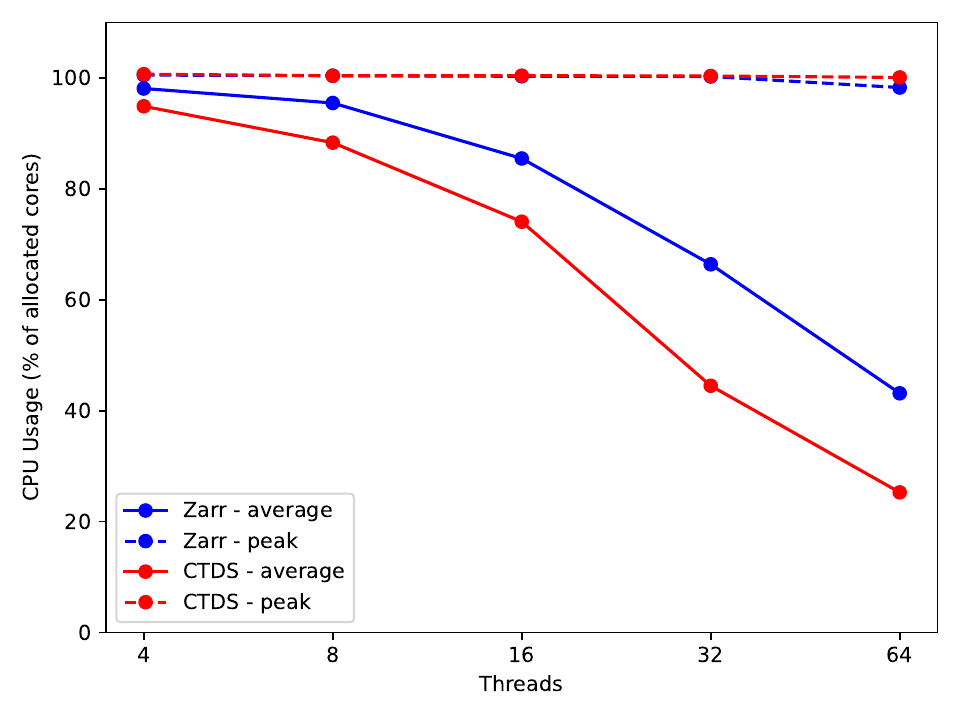}
        \caption{Solving for residual delays and phase offsets (\textbf{K}).}
    \end{subfigure}%
    ~ 
    \begin{subfigure}[t]{\columnwidth}
        \centering\captionsetup{width=.8\linewidth}
        \includegraphics[width=\columnwidth]{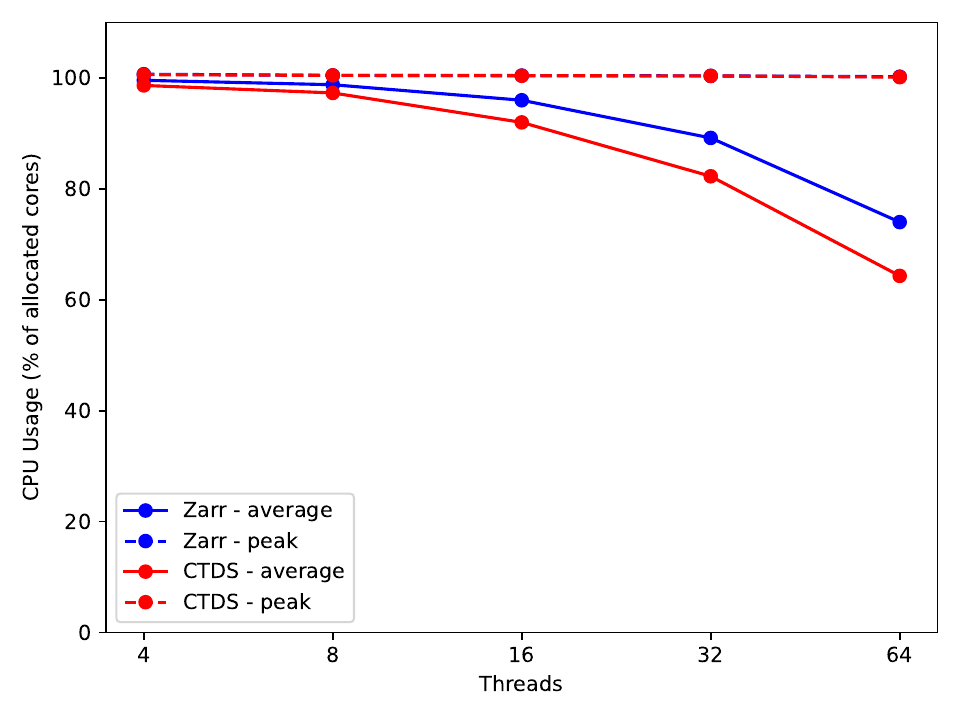}
        \caption{Solving for residual delays and phase offsets (\textbf{K}) and a direction-dependent, complex-valued gain (\textbf{dE}).}
    \end{subfigure}
    \caption{Benchmarking results comparing \quartical{} run on a Measurement Set backed by the CTDS (\textit{left}) and \quartical{} run on a Measurement Set backed by \zarr{} (\textit{right}). \textit{Top row:} Wall time in minutes as a function of allocated threads. \textit{Middle row:} Memory footprint in gigabytes as a function of allocated threads. \textit{Bottom row:} CPU utilization as a function of allocated threads.}
    \label{fig:ctds_vs_zarr}
\end{figure*}

Another interesting topic which was alluded to in $\S$~\ref{subsection:ingest} is the difference in performance characteristics between Measurement Sets backed by \zarr{} and those backed by the CTDS. To test this behaviour (as well as validate that \quartical{} functions transparently with both), we utilized \daskms{} to convert the entire CTDS dataset described in the preceding experiments into one backed by \zarr{}. \quartical{} was then run on both datasets using the threaded scheduler and identical settings. The results appear in Figure \ref{fig:ctds_vs_zarr}.

The wall times are approximately as one would expect; \quartical{} is faster when using a \zarr{}-backed Measurement Set than one backed by the CTDS. This is due to the fact that reading and writing to the \zarr{}-backed Measurement Set does not require the GIL and consequently doesn't block the execution of other threads. Note that the arguments from $\S$~\ref{subsection:qcvscc} regarding the deterioration of linear scaling at high thread counts still apply. In the \textbf{KdE} case, the problem is more compute-bound and the discrepancy between the \zarr{} and CTDS approaches is less obvious as the I/O is a much smaller fraction of the overall run time.

The memory results for this comparison are particularly interesting as the \zarr{} experiment uses more memory than the CTDS experiment for all thread counts. The reason for this is somewhat subtle but is related to the fact that the slower data access in the CTDS essentially starves \quartical{} of data, i.e. the \zarr{} case uses more memory as it is not constrained by the GIL and can more readily read the data into memory. We note that in the \textbf{KdE} case with many threads, the memory footprints tend towards the same number - this is once again because more time is spent doing compute than I/O and the overheads are effectively averaged out. 

Unlike the previous experiment, the peak and average CPU usage plots are included as they are interesting and easy to understand. These plots support our previous claims as the \zarr{} results indicate consistently higher average CPU usage than the CTDS case. This is because there is less interaction with the GIL and the reads are typically faster. Additionally, some CPU usage will be associated with the decompression of the \zarr{} data, which is compressed automatically during conversion from the original Measurement Set. On the whole, CPU usage is poor for the \textbf{K}-only case. This is due to the fact that the problem is dominated by I/O and the threads are starved of work. The CPU usage improves substantially for the \textbf{KdE} chain as we move from I/O-dominated to a more compute-dominated regime. The reason for the less than 100\% average efficiency is again due to the I/O component, during which many CPUs may be sitting idle due to finite disk throughput. As mentioned in $\S$~\ref{subsection:ingest}, interleaving compute and I/O is one of the areas in which \quartical{} may be improved in the future.

\subsection{Profiling: \quartical{} on AWS}
\label{subsection:qconaws}

In addition to the preceding experiments which were run on hardware under our control, it was necessary to validate that \quartical{} can function in a truly distributed environment. While \quartical{} has been run on a supercomputer in the past, limited access to supercomputing resources motivated us to explore using the cloud. These experiments make use of AWS\footnote{\url{https://aws.amazon.com/}} (Amazon Web Services) but it is important to note that \quartical{} is unaware of the specific cloud provider in use; \quartical{} should be compatible with any service on which a \dask{} scheduler and worker nodes can be instantiated. The infrastructure necessary to run \quartical{} in this fashion is handled by \stimela{}, and an in-depth description of this functionality is deferred to Paper IV. Suffice it to say that \stimela{} has the necessary components to communicate with a Kubernetes\footnote{\url{https://kubernetes.io/}} \citep{brewer2015} cluster running on AWS and ultimately bring up a scheduler node, a runner node, and a number of compute nodes on which \quartical{} can be run.

For the first experiment, we consider deriving 1GC solutions for the data which will be used in Paper III. The full observational details will be left for that paper, but the calibrator observation in question is an L-band, 4096 channel, 8 second integration MeerKAT observation of J1939-6342. This data was manually flagged prior to calibration and contains approximately 103GB of visibilities in its DATA column. The goal is to solve for a \textbf{GKB} Jones chain, where \textbf{G} is a diagonal, complex-valued gain that captures the mean amplitude and phase error per scan, \textbf{K} is a residual delay term, and \textbf{B} is the bandpass; a diagonal, complex-valued gain that models the antennas' response as a function of frequency. For the purposes of this experiment, these values are computed per scan. Interpolation of the gain solutions from the calibrator to the target is not included in these experiments.

Solving a Jones chain of this type is typically a difficult problem as computing the residual delay solutions requires the entirety of the frequency axis to be in memory whereas computing the bandpass requires the entirety of the time axis to be in memory. As such, we are in a regime with very large chunks and have limited data partitions over which we can parallelize. Fortunately, as discussed in $\S$~\ref{subsection:solvers}, it is possible to use \quartical{}'s nested parallelism model to leverage the available hardware.

Amazon EC2\footnote{\url{https://aws.amazon.com/ec2/}} (Elastic Compute Cloud) offers a staggering number of instance (node) types with a plethora of different features. In this case, we decided to make use of c6in.8xlarge instances\footnote{\url{https://aws.amazon.com/ec2/instance-types/c6i/}} which boast 16 vCPU (8 physical cores, 16 threads with simultaneous multithreading) with up to 3.5GHz clock speeds, 64 GiB of RAM, and up to 50 Gbps network bandwidth. It is this last property that motivated our choice as the data for this experiment was stored as a \zarr{}-backed Measurement Set stored on Amazon S3 (a cloud-based object store). The throughput of S3 scales linearly with the number of nodes and, given a sufficient number of nodes, it is possible to reach very high throughput. This is necessary to ensure that the problem is not entirely (network) I/O bound. We note in passing that, while we have elected not to present the results here, we did experiment with a variety of different cloud-based storage options including Amazon EFS (Elastic File System) and Amazon EBS (Elastic Block Store). Neither of these can compete with the throughput of S3 in the distributed case.

\begin{figure}
    \centering
    \includegraphics[width=\columnwidth]{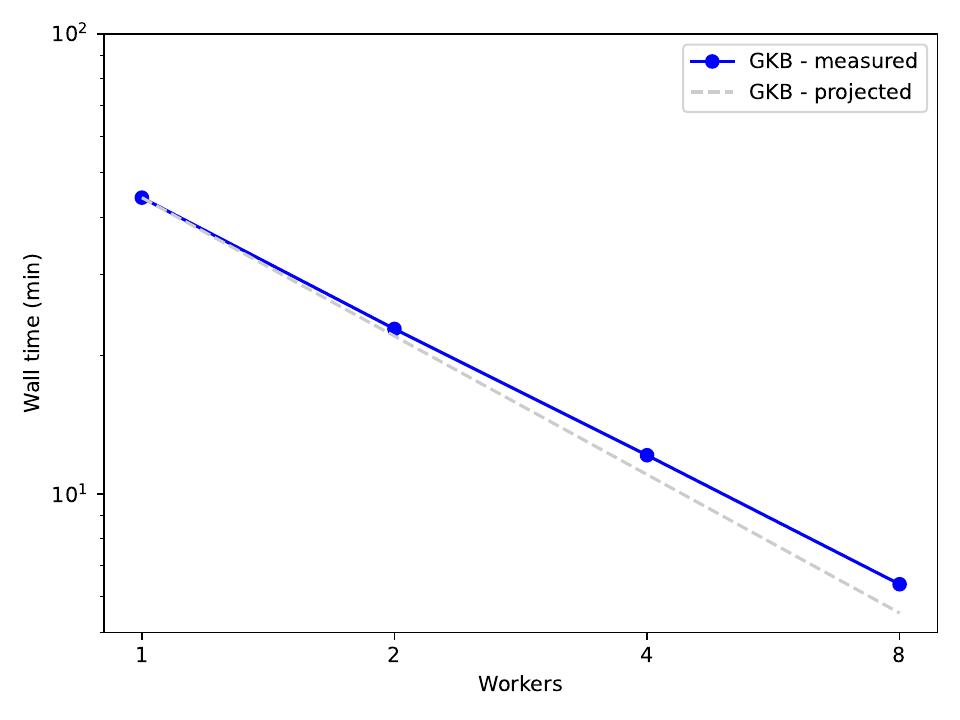}
    \caption{Wall time in minutes as a function of the number of \dask{} workers when using \quartical{} to solve for a diagonal, complex-valued gain (\textbf{G}), a residual delay (\textbf{K}) and a bandpass (\textbf{B}) on a per-scan basis. The projected curve is generated by presuming linear scaling and extrapolating from the measured value for the single worker case.}
    \label{fig:aws_large_chunks}
\end{figure}

We calibrated the data several times using an increasing number of cluster nodes and recorded the wall time as a function of cluster size. In this instance, as there were eight calibrator scans in the Measurement Set, there were only eight data partitions (containing a single chunk each) to be processed in parallel. Consequently, we could only scale up to a maximum of eight nodes (\dask{} workers), as any additional resources would be largely unused. Each \dask{} worker was configured to use a single \dask{} thread, but the \numba{}-based parallelism at the solver level was configured to use the full 16 threads on each node. This experiment made use of the \textit{AutoRestrictor} plugin mentioned in $\S$~\ref{subsection:graph_execution} which coerces all tasks associated with a given data partition to be computed on a specific node. This means that the experiment is free from unnecessary data transfers, which can have catastrophic consequences when dealing with very large chunks of data. The results appear in Figure \ref{fig:aws_large_chunks}.

The results clearly indicate that \quartical{} scales nearly linearly as a function of worker nodes. This is precisely the desired behaviour. The small deviation from linear scaling can be attributed to our previous observation that this is a real observation and an iterative algorithm - in the case where there is only a single chunk to process per node, the wall time will be determined by the slowest chunk. No attempt was made to characterize \quartical{}'s memory behaviour for the cloud-based experiments as this information is not readily available. However, we can say with certainty that it remained within the bounds set by the EC2 instance types as failure to do so would have resulted in errors. 

The second experiment aims to establish \quartical{}'s performance in the cloud for a self-calibration problem. Self-calibration typically allows for the use of smaller chunks as the solution intervals do not need to be as large as in the previously described 1GC example. These smaller chunks mean that there are far more opportunities for \dask-based parallelism.

\begin{figure}
    \centering
    \includegraphics[width=\columnwidth]{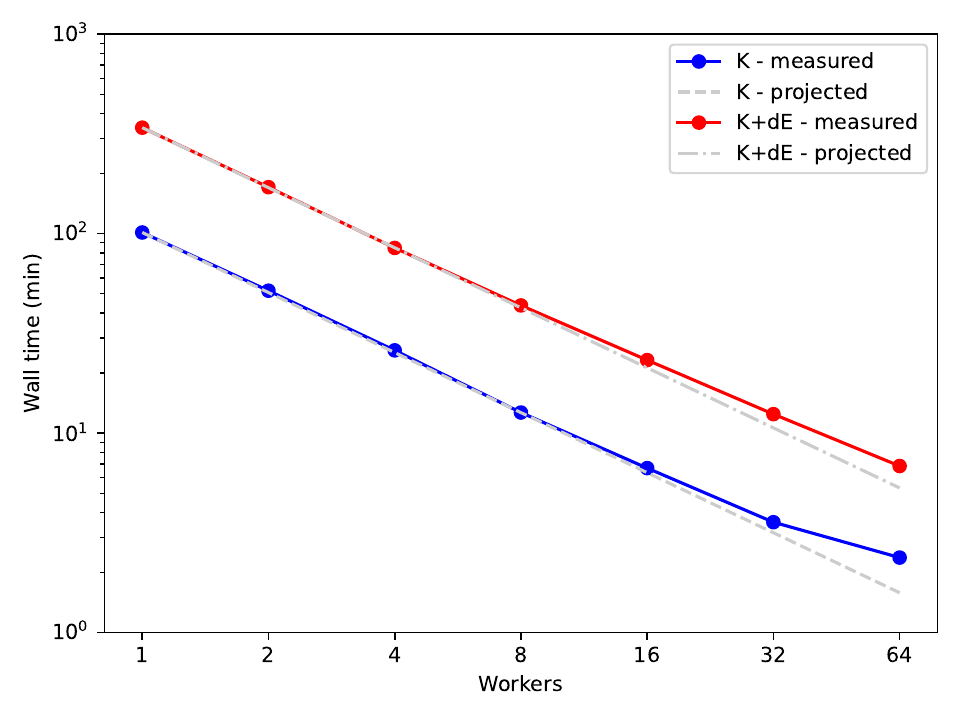}
    \caption{Wall time in minutes as a function of the number of \dask{} workers when using \quartical{} to solve for residual delays and phase offsets (\textbf{K}) and a Jones chain consisting of \textbf{K} and a direction-dependent, diagonal, complex-valued gain (\textbf{dE}).}
    \label{fig:aws_small_chunks}
\end{figure}

We made use of the data described in $\S$~\ref{subsection:qualitative_results}, converted to \zarr, and uploaded to Amazon S3. The gain calibration is the same as that described in $\S$~\ref{subsection:qcvscc}, including identical chunking but without producing any visibility outputs i.e. the only data product is the gains. As the chunks were smaller, we elected to use c6in.2xlarge instances. These instances have 8 vCPU (4 physical cores, 8 threads with simultaneous multithreading) with up to 3.5GHz clock speeds, 16 GiB of RAM, and up to 40 Gbps network bandwidth. These are relatively modest instances and cost roughly 0.6 USD per hour to run. Due to the limited memory available on these nodes, only a single \dask{} thread was used per worker, but all 8 threads were used by the \numba{}-based parallelism in the solvers. We recorded and plotted the wall time as a function of cluster size (number of \dask{} workers) in Figure \ref{fig:aws_small_chunks}.

Once again, there are strong indications of the desired linear scaling. This is seen in both the \textbf{K} and \textbf{KdE} results. Both sets of results begin to deviate from this behaviour at high worker counts for the same reasons as described in $\S$~\ref{subsection:qcvscc}; at 64 threads, there are only 3.7 chunks of data to process per worker. This means that our wall time measurements will typically be biased by the slowest running chunks. It is likely that for bigger problems with a larger number of chunks that this deviation would disappear (for these worker counts). No results were generated beyond 64 workers as requisitioning 128 workers on EC2 is not possible by default.

This experiment made use of the \textit{SolverRestrictor} scheduler plugin described in $\S$~\ref{subsection:graph_execution}, as we did not want to limit the number of \dask{} workers we could use to the number of data partitions as was done for the preceding experiment. This means that these results will include some data transfers and overhead, but this does not seem to have impacted the results in a serious manner. Attempts to run these experiments with the scheduler plugin disabled resulted in instability and out-of-memory errors, highlighting one of \dask{}'s primary limitations.

\section{Conclusions}

We have presented \quartical{}, a new Python package for the calibration of radio interferometer data. \quartical{} is the successor to \cubical{} and has its roots in the same complex optimization framework. That framework has been further simplified and extended to include Jones chains containing a mixture of parameterized and non-parameterized terms.

\quartical{} makes use of the \textsc{Africanus} ecosystem described in this paper series to tackle large calibration problems. \daskms{} provides an interface to the Measurement Set and allows \quartical{} to use \dask{} to parallelize (and ultimately distribute) the problem. \numba{} is used to optimize performance critical sections of the code, and \xarray{} and \zarr{} provide a convenient on-disk representation for the gain solutions. \stimela{} integration makes including \quartical{} in new data reduction pipelines trivial.    

Our results show that \quartical{} is successful in performing calibration from 1GC through to 3GC, which is not typically the case for existing software. Beyond simply calibrating data, we have demonstrated that \quartical{} convincingly outperforms \cubical{} in terms of both memory footprint and wall time. The results also indicate that \quartical{} is capable of running and scaling on a variety of hardware, from a user's laptop, to a single large compute node, and ultimately a distributed cluster running in the cloud. This last case is particularly compelling as it suggests that \quartical{} is capable of calibrating incredibly large problems in a reasonable amount of time given adequate resources.  

Whilst developing \quartical{}, we have become acutely aware of the limitations of \dask{} and have developed tools to circumvent them. \quartical{} makes use of scheduler plugins to manipulate task placement in distributed environments to avoid data transfer and stabilize memory usage. Additionally, the functional style \dask{} enforces can lead to excess memory use which we are in the process of addressing.

\quartical{} already boasts a wide variety of features but there are still many ways in which in could be improved. Chief among them is tighter integration with \pfb{} in order to generate model visibilities on the fly by degridding \pfb{} models. This functionality is already in development. There are also substantial gains to be made in writing GPU versions of the solvers.

Finally, we stress that \quartical{} is already publicly available and in need of additional users.

\section*{Acknowledgements}

Funding: OMS's and JSK's research is supported by the South African Research Chairs Initiative of the Department of Science and Technology and National Research Foundation (grant No. 81737).

The MeerKAT telescope is operated by the South African Radio Astronomy Observatory, which is a facility of the National Research Foundation, an agency of the Department of Science and Innovation. 

We would like to thank Ross Davies, Ilze van Tonder, Christopher Voges, Peter Fosseus, Nawaal Adams, Dietrich Keller and their colleagues at Silicon Overdrive, as well as Agnat Max Makgoale (Amazon Web Services) for invaluable technical and logistical assistance over the course of this work.

\section*{Data Availability}
The MeerKAT data used for the benchmarks in this paper is publicly available via the SARAO archive\footnote{\url{https://archive.sarao.ac.za}}, under proposal IDs SSV-20200715-SA-01 and SCI-20190418-SM-01.

\appendix

% \input{appendices/appendix1.tex}

\bibliographystyle{elsarticle-harv} 
\bibliography{quartical}



%% For example:

%% \bibitem[Aladro et al.(2015)]{Aladro15} Aladro, R., Martín, S., Riquelme, D., et al. 2015, \aas, 579, A101

%%\end{thebibliography}

\end{document}